\documentclass[fleqn,twoside,twocolumn,nofootinbib]{revtex4} % Specifies the document class %,unsortedaddress
\usepackage[sec]{ujp} % \usepackage[]{ujp} for cyrillic, \usepackage[web]{ujp} for web
\begin{document}
\title[INTERACTION BETWEEN AN ISOTROPIC NANOPARTICLE]%колонтитул
{INTERACTION~~\, BETWEEN~~\, AN~~\, ISOTROPIC\\ NANOPARTICLE AND
DRIFTING ELECTRONS\\ IN A
QUANTUM WELL}%
\author{V.A.~Kochelap }%1 автор
\affiliation{V. Lashkaryov Institute of Semiconductor Physics, Nat. Acad. of Sci. of Ukraine}%институт
\address{41, Prosp. Nauky, Kyiv 03680, Ukraine}%адрес
\email{kukhtaruk@gmail.com}%e-mail
\author{S.M. Kukhtaruk}%
\affiliation{V. Lashkaryov Institute of Semiconductor Physics, Nat. Acad. of Sci. of Ukraine}%институт
\address{41, Prosp. Nauky, Kyiv 03680, Ukraine}%адрес
\email{kukhtaruk@gmail.com}

\udk{53.01, 53.04} \pacs{72.30.+q, 73.20.Mf,\\[-3pt] 73.63.Hs, 73.63.Kv.}
\razd{\secviii}
\setcounter{page}{367}%
\maketitle

%\makeatletter
%\renewcommand{\thesection}{\arabic{section}}
%\renewcommand{\p@subsection}{}
%\renewcommand{\thesubsection}{\arabic{section}.\arabic{subsection}}
%\renewcommand{\p@subsubsection}{}
%\renewcommand{\thesubsubsection}
%{\arabic{section}.\arabic{subsection}.\arabic{subsubsection}}
%\makeatother

\begin{abstract}
A hybrid system composed of an isotropic nanoparticle and a
semiconductor heterostructure with a quantum well has been
considered. The nanoparticle is supposed to be polarizable in an
external electric field. A theoretical model of the hybrid system is
substantiated and formulated. Exact solutions of the model equations
are obtained. The frequencies of charge oscillations in the hybrid
system and their damping owing to the dipole--plasmon interaction
are found, the damping mechanism being similar to that of Landau
damping. The space-time behavior of concentration perturbations in
the two-dimensional electron gas is analyzed, and the polarization
oscillations of a nanoparticle are studied. The induced polarization
of a nanoparticle at nonzero electron drift velocities is found to
have a complicated dynamics. In particular, the polarization vector
circulates along elliptic trajectories for two of three frequency
dispersion branches. If the electric current flows through the
quantum well due to an applied electric field, the damping of
oscillations in the hybrid system is replaced by their growth in
time, which corresponds to the electric instability of the system.
New phenomena in hybrid systems can be used to excite the emission
of nanoparticles by an electric current and to electrically
stimulate the emission in the terahertz spectral range.
\end{abstract}

\section{Introduction}

Physical phenomena in the terahertz (THz) spectral range are intensively
studied, which is associated with the fundamental importance of new terahertz
physics, as well as with a considerable number of feasible applications. In
particular, many researches aim at constructing new sources and
detectors of THz radiation.

In this context, of great interest are the researches of
semiconductor heterostructures with quantum wells (QWs), in which
collective oscillations of a two-dimensional electron gas (2DEG),
i.e. plasmons, can be excited. The frequencies of plasmon
oscillations belong to the THz spectral range. In a homogeneous
2DEG, plasmons are stable, i.e. their oscillations attenuate in time
owing to the electron scattering by various crystal defects and by
means of the known mechanism of Landau damping. Various methods were
proposed in order to achieve the instability and obtain
oscillation-growth effects intended to be used for the amplification
and the generation of THz radiation. Among those, a possibility of
the instability excitation owing to the electron drift in an
electric field \cite{rev-1, rev-2, Wilkins}, various variants of
two-beam instability \cite{Wilkins, Gribnikov}, and others were
examined. However, for such instabilities to be realized in a
spatially uniform plasma, very high drift velocities, which are
difficult to be obtained experimentally, are required. The situation
essentially changes in spatially non-uniform or finite systems
\cite{f-length-1}. For example, as was shown in works
\cite{Dyakonov}, if special types of contacts are applied to a
quantum well with a finite length, the electron system becomes
unstable already at moderate drift velocities of electrons. This
instability has been studied experimentally \cite{Knap-2004,
Knap-2005, Knap-2008}.\looseness=1

Another class of objects, which are active in the THz spectral range, includes
quantum dots \cite{Demel_1990, old-QDs-1, old-QDs-2, old-QDs-3}, molecules and
some molecular compounds \cite{Yu_2004, Maistrenko_1999, Balu_1999}, shallow
impurity centers \cite{Burghoon_1994, Kalkman_1996, Allen_2005}, and so forth.
For brevity, let us call such \textquotedblleft
zero-dimensional\textquotedblright\ objects as nanoparticles (NPs).

Hybrid systems consisting of nanoparticles and heterostructures with
free electrons constitute a new type of \textit{heterodimensional}
objects, which should demonstrate essentially new properties and
effects. In particular, if the oscillation frequencies of
zero-dimensional NPs and two-dimensional plasmons are in the THz
spectral range, hybrid systems may reveal new properties at
extremely high frequencies. Therefore, the study of the interaction
between such NPs and plasmons under equilibrium and nonequilibrium
conditions seems well-timed.

%Fig. 1
\begin{figure}% figure* for wide figure, [h] [!] to change the placement
\includegraphics[width=\column]{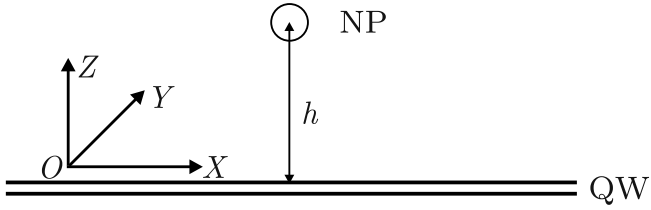}
\vskip-3mm\caption{Schematic drawing of the hybrid system. The $OY$
axis is perpendicular to the figure plane }\label{Fig-1}
\end{figure}

In work \cite{Kukhtaruk_UJP}, the studies of the interaction between a remote
nanoparticle and a drifting 2DEG at relatively low frequencies, i.e. when the
application of the \textquotedblleft low-frequency\textquotedblright%
\ drift-diffusion approximation is valid, were started. In this work, we
present the results of our researches of THz-frequency properties of hybrid
systems in the \textquotedblleft high-frequency\textquotedblright%
\ approximation for the electron motion in a 2DEG. We revealed
collective oscillations of interacting nanoparticles and the 2DEG.
In particular, we determined the frequencies of collective
oscillations and their extra damping stemming from the interaction
between the nanoparticle and plasmons. The damping character is
similar to that of Landau damping. This system was demonstrated to
become unstable under definite nonequilibrium conditions, i.e. there
emerges an increase of collective oscillations in time induced by an
electric current. The nanoparticle destroys the translational
symmetry, and the system becomes non-uniform, which promotes the
appearance of an instability, similarly to what takes place in
non-uniform or spatially confined
plasma~\cite{f-length-1}.\looseness=1

The structure of the paper is as follows. In Section~2, the model of
a hybrid system is discussed, and the basic equations are
substantiated. In Section~3, the exact solutions of the equations
are obtained, and their general properties are examined. In
Section~4, the main attention is given to the stability/instability
of a hybrid system under the condition of two-dimensional electron
drift, the numerical results are presented, and the specific
examples of nanoparticles and heterostructures with two-dimensional
electrons are considered. The features of coordinated space-time
charge waves are dealt in Section~5. The properties of the NP dipole
moment, plasmon field, and emission by the hybrid system are studied
in Section~6. Conclusions are presented in Section~7.\looseness=1

\section{Model and Basic Equations}

Consider a hybrid system represented schematically in Fig.~1. It consists of a
spherical nanoparticle, which can be polarized in an external field, and the
heterostructure with a quantum well, in which two-dimensional electrons are
localized. The distance between the nanoparticle and the quantum well is $h$.
The electrons and the nanoparticle interact by means of the electrostatic
field $\mathbf{E}=-\mathbf{\nabla}\phi$, where $\phi(x,y,z,t)$ is a
self-consistent electrostatic potential. The latter is described by the
Poisson equation
%1
\begin{equation}
\frac{\kappa}{4\pi}\Delta\phi=e({n}-n_{0})\delta(z)+\left( {\
\mathbf{D}}(t)\mathbf{\nabla}\right)  \delta(x)\delta(y)\delta
(z-h),\label{Poisson}%
\end{equation}
where $\kappa$ is the dielectric permittivity, $e$ the electron
charge, ${n}(x,y,t)$ the surface concentration of electrons, $n_{0}$
the equilibrium surface concentration of electrons, $\mathbf{D}(t)$
the electric dipole moment of a nanoparticle, and $\delta(x)$ the
Dirac delta-function.

It is convenient to express the scalar potential as a sum of two terms
associated with the electrons and the NP, ${\phi(x,y,z,t)=\phi_{e}%
(x,y,z,t)+\phi_{d}(x,y,z,t)}$, where $\phi_{e}$ and $\phi_{d}$ are determined
from the system of equations
%2
\begin{equation}%
\begin{cases}
\displaystyle{\kappa\Delta\phi_{e}=4\pi e({n}-n_{0})\delta
(z),}\cr\displaystyle{\kappa\Delta\phi_{d}=4\pi\left(  {\mathbf{D}%
}(t)\mathbf{\nabla}\right)  \delta(x)\delta(y)\delta(z-h).}%
\end{cases}
\label{2xPoisson}%
\end{equation}

To describe electrons in the quantum well (the plane $z=0$), we use
Euler's equation of motion and the continuity equation, which couple
the surface concentration ${n}(x,y,t)$ and the electron velocity
$\mathbf{\mathfrak{v}}(x,y,t)$ with the electrostatic field. In the
general case, we assume that electrons move with an average drift
velocity $\mathbf{v_{0}}$. Then, the equations are \cite{Dyakonov}
%3
\begin{equation}%
\begin{cases}
\displaystyle{\frac{\partial\mathbf{\mathfrak{v}}}{\partial
t}+\left(
{\mathbf{\mathfrak{v}}}{\mathbf{\nabla}}_{\parallel}\right)
{\mathbf{\mathfrak{v}}}=\left.
\frac{e}{m}\mathbf{\nabla}\phi\right\vert
_{z=0}-\frac{\mathbf{\mathfrak{v}-v_{0}}}{\tau_{p}}}\,,\cr\displaystyle{\frac
{\partial{n}}{\partial t}+{\mathbf{\nabla}}_{\parallel}\left(
\mathbf{\mathfrak{v}}{n}\right)  =0}\,,
\end{cases}
\label{hydrodynamics}%
\end{equation}
where $m^{\ast}$ is the effective mass of an electron, the term ${\frac
{\mathbf{\mathfrak{v}-v_{0}}}{\tau_{p}}}$ makes allowance for the electron
scattering by crystal defects, and $\tau_{p}$ is the time of electron momentum
relaxation. The subscript $\parallel$ is used for quantities and differential
operators dealing with the plane $z=0$ only. For instance, the Laplace
operator in the plane $z=0$ is ${\Delta_{\parallel}=\frac{\partial^{2}%
}{\partial x^{2}}+\frac{\partial^{2}}{\partial y^{2}}}$. At the same time, the
Laplace operator in the volume is ${\Delta=\Delta_{\parallel}+\frac
{\partial^{2}}{\partial z^{2}}}$. We will also mark the real and imaginary
parts of complex numbers as primed quantities, e.g., ${Z=Z^{\prime}%
+iZ^{\prime\prime}}$.

In the two-dimensional electron gas, there emerge collective oscillations of
the charge density, which are called plasmons. The hydrodynamic model with
Eqs.~(\ref{hydrodynamics}) is valid, when the frequency of these oscillations,
$\omega$, and the absolute value of the wave vector, $k$, satisfy the
conditions
%4
\begin{equation}
\omega\tau_{p}\gg1,\,\,ql_{p}\sim l_{p}/h\gg1\,,\label{crit}%
\end{equation}
where $l_{p}$ is the mean free path of electrons. The formulation of the
problem includes only one characteristic length scale; it is the distance $h$
between the plane of the 2DEG and the NP. Therefore, one may expect that the wave
vectors of plasmons, which make the major contribution to the interaction
between the electrons and the NP, should be of the order of $1/h$. Note that
inequalities (\ref{crit}) correspond to the ballistic character of the electron
motion in the spatial region, which is actual for the nanoparticle--plasmon interaction.

Suppose that the velocity of electrons is ${\mathbf{\mathfrak{v}%
}(x,y,t)=\mathbf{v_{0}}+\mathbf{v}(x,y)e^{-i\omega t}},$ and their
concentration ${{n}(x,y,t)=n_{0}+n(x,y)e^{-i\omega t}}$, with ${v\ll
v_{0}}$ and ${n\ll n_{0}}$. Let the time dependences of potentials
also vary harmonically, i.e.
${\phi_{e}(x,y,z,t)=\varphi_{e}(x,y,z)e^{-i\omega t}}$ and
${\phi_{d}(x,y,z,t)=\varphi_{d}(x,y,z)e^{-i\omega t}}$. The
frequency $\omega$ in these formulas should be determined by solving
the equations; generally speaking, it is a complex-valued quantity,
which corresponds to the oscillations (damped or growing in time) in
the hybrid system. (If ${\omega^{\prime\prime}>0}$, the system is
unstable, and ${\omega ^{\prime\prime}}$ is called the instability
increment.) The quantities $\mathbf{v}(x,y)$ and $n(x,y)$ describe
the spatial dependences of electron velocity and concentration
perturbations, respectively.

According to Fig.~1, the nanoparticle is located at the point $\left(
x=0,y=0,z=h\right)$. In an external electric field, it becomes polarized and
can be characterized by an electric dipole moment, which is determined by the
relation
%5
\[
 {\mathbf{D}(t)= \mathbf{d}(\omega)e^{-i\omega
t},}
\]
\begin{equation}
{\mathbf{d}(\omega)=-\left.\beta_0(\omega) \mathbf{\nabla}
\varphi_e\right|_{x=y=0,z=h},} \label{dipole}
\end{equation}
where $\beta_{0}(\omega)$ is the nanoparticle polarizability. Note
that formula (\ref{dipole}) includes only the electron potential
and, hence, the dipole self-action is excluded from consideration.
If the nanoparticle is isotropic, its polarizability can be
expressed in the standard form \cite{Davydov}
%6
\begin{equation}
\displaystyle{\beta_{0}(\omega)=-\sum\limits_{f}\frac{2\omega_{f}e^{\ast^{2}}%
}{\hbar}\,\frac{|\langle f|x|0\rangle|^{2}}{\omega^{2}-\omega_{f}^{2}%
+i\omega/\tau_{f}}}\,, \label{beta0main}%
\end{equation}
where $\hbar$ is the reduced Planck constant, $\hbar\omega_{f}$ is
the energy of the $f$-th nanoparticle level, $\tau_{f}$ is the
lifetime of this level, $\langle f|x|0\rangle$ is the matrix element
of the dipole transition between the ground and $f$-th excited
states.

The largest contribution to polarizability (\ref{beta0main} ) is
given by the transition between the ground and the first excited
state. In this case, the polarization is
%7
\begin{equation}
\displaystyle{\beta_{0}(\omega)=-\frac{e^{\ast^{2}}}{\hbar}\,\frac
{|\langle1|x|0\rangle|^{2}}{\omega-\omega_{0}+i\gamma_{0}}}\,,\label{beta0}%
\end{equation}
where the characteristic frequency of dipole oscillations, ${\omega_{0}}$, and
the damping of dipole oscillations corresponding to the transition between the
ground and the first excited state, ${\gamma_{0}=1/(2\tau_{1})}${, }are
introduced. While deriving formula (\ref{beta0}), an assumption was made that
$\omega$ is close to $\omega_{0}$, because the main effects are expected to
take place at frequencies close to the characteristic frequency of dipole oscillations.

Equations (\ref{Poisson})--(\ref{hydrodynamics}), (\ref{dipole}),
and (\ref{beta0}) compose a basic system for the problem under
consideration.

\section{Solutions of the Equations and Their General Properties}

The dipole potential, being the solution of the second Poisson equation in
system (\ref{2xPoisson}), has the well-known form
%8
\begin{equation}
\displaystyle{\varphi_{d}=\frac{1}{\kappa}\frac{d_{x}(\omega)x+d_{y}%
(\omega)y+d_{z}(\omega)(z-h)}{(x^{2}+y^{2}+(z-h)^{2})^{\frac{3}{2}}}},
\label{phi_d}%
\end{equation}
where $d_{x}$, $d_{y}$, and $d_{z}$ are the components of the dipole moment
$\mathbf{d}(\omega)$. Below, for all quantities depending on coordinates, the
two-dimensional Fourier transformation is used:
%9
\begin{equation}
\displaystyle{\varphi_e(\mathbf{r},z)=\!\int\!  d^2\!k\,
\varphi_k(z)\, e^{i\mathbf{k}\mathbf{r}}},
 \label{phi_e}
 \end{equation}
 %10
 \begin{equation}
\displaystyle{n(\mathbf{r})=\!\int\! d^2\!k\, n_k\,
e^{i\mathbf{k}\mathbf{r}}\,,\,\,\,\,\, \psi(\mathbf{r})=\!\int\!
d^2\!k\, \psi_k\, e^{i\mathbf{k}\mathbf{r}},} \label{n_e}
\end{equation}
where ${\mathbf{r}=(x,y)}$, ${\mathbf{k}=(k_{x},k_{y})}$, and ${\psi
(\mathbf{r})=\left.  \varphi_{d}(\mathbf{r},z)\right\vert _{z=0}}$.
The quantities $\varphi_{k}(z)$, $n_{k}$, and $\psi_{k}$ are the
Fourier transforms of the electron potential, surface concentration,
and potential induced by the dipole in the quantum-well plane
($z=0$), respectively. Then, the Poisson equation for
$\varphi_{k}(z)$ and the required boundary conditions are
%11
\begin{equation}
\begin{cases}
\displaystyle{\frac{d^2 \varphi_k^{\pm}}{d z^2} - k^2
\varphi_k^{\pm} =0},\\[2mm]
\displaystyle{\left.\varphi_k^{\pm}\right|_{z\rightarrow
\pm\infty}\rightarrow 0,}\\[2mm]
\displaystyle{\left.\varphi_k^+\right|_{z=0+\varepsilon}=\left.\varphi_k^-\right|_{z=0-\varepsilon},}\\[2mm]
\displaystyle{\left.\frac{d \varphi_k^+}{d z}
\right|_{z=0+\varepsilon}-\left.\frac{d \varphi_k^-}{d
z}\right|_{z=0-\varepsilon}=\frac{4\pi e
n_k}{\kappa},}\label{kpoisson}
\end{cases}
\end{equation}
where $\varphi_{k}^{+}(z)=\varphi_{k}(z)$ at $z\geq0$ and $\varphi_{k}%
^{-}(z)=\varphi_{k}(z)$ at $z\leq0$, $k=\sqrt{k_{x}^{2}+k_{y}^{2}}$, and
$\varepsilon\rightarrow+0$.

The solution of system (\ref{kpoisson}) and the functions $\psi_{k}$
and $n_{k}$ are
%12
\begin{equation}
\begin{cases}
\displaystyle{\varphi_k=-\frac{2\pi e n_k}{\kappa k}e^{-k |z|}}\\[2mm]
\displaystyle{\psi_k=-\frac{1}{2\pi\kappa}\left(i\frac{d_xk_x+d_yk_y}{k}
+ d_z\right)e^{-kh}}\\[3mm] \displaystyle{n_k=-\frac{e n_0
k^2}{m}\frac{\left.
\varphi_k\right|_{z=0}+\psi_k}{(\omega-\mathbf{v_0
k})(\omega-\mathbf{v_0 k}+i/\tau_p)}}.\label{3eq}
\end{cases}
\end{equation}
Note that, in the limiting case where the dipole is infinitely far from the
2DEG (i.e. ${h\rightarrow\infty}$), solutions (\ref{3eq}) easily bring about
the dispersion law for collective excitations in the drifting 2DEG, i.e.
drifting plasmons,
%13
\begin{equation}
\displaystyle{\omega_{\pm}=\mathbf{v_{0}k}\pm\sqrt{\frac{2\pi e^{2}n_{0}%
}{\kappa m}k-\frac{\gamma_{p}^{2}}{4}}-\frac{i\gamma_{p}}{2},} \label{plasma}%
\end{equation}
where the notation ${\gamma_{p}=1/\tau_{p}}$ is used. On the other hand, the
zero of the denominator of $\beta_{0}(\omega)$ (Eq.~(\ref{beta0})) corresponds
to the frequency and the damping of a dipole, $\omega=\omega_{0}%
-i\gamma_{0}$.

If the distance $h$ is finite, system (\ref{3eq}) and relations (\ref{dipole})
and (\ref{beta0}) can be used to derive an integral equation, e.g., for the
Fourier transform of the concentration of electrons interacting with the
dipole,
%14
\begin{equation}
\displaystyle{n_{k}=-\frac{e^{2}n_{0}}{\kappa^{2}m}\frac{\beta_{0}%
(\omega)ke^{-kh}(k_{x}I_{x}+k_{y}I_{y}+kI_{z})}{(\omega-\mathbf{v_{0}%
}\mathbf{k})(\omega-\mathbf{v_{0}}\mathbf{k}+i\gamma_{p})-\frac{2\pi
e^{2}n_{0}k}{\kappa m}}},\label{integal equation}%
\end{equation}
where ${I_{x}}$,$\ {I_{y}}$,\ and ${I_{z}}$ are the functionals of $n_{k}$,%
\[
\displaystyle{I_{x}\!=\!\!\!\int\!\!d^{2}q\frac{q_{x}}{q}n_{q}e^{-qh}%
},\,\displaystyle{I_{y}\!=\!\!\!\int\!\!d^{2}q\frac{q_{y}}{q}n_{q}e^{-qh}%
},\displaystyle{I_{z}\!=\!\!\!\int\!\!d^{2}qn_{q}e^{-qh}}.
\]
With the notations
\[
\displaystyle {B_0=-\frac{e^2 n_0}{\kappa^2 m}\beta_{0}(\omega)},
\]
\[
\displaystyle{\Delta_e(\omega,k)=(\omega-\mathbf{v}_0
\mathbf{k})(\omega-\mathbf{v}_0 \mathbf{k}+i \gamma_p)-\frac{2 \pi
e^2n_0k}{\kappa m}},
\]
the integral equation (\ref{integal equation}) yields the
following system of algebraic equations:
%15
\begin{equation}%
\begin{cases}
\displaystyle{I_{x}=B_{0}(S_{x}I_{x}+S_{0}I_{z}),}\cr\displaystyle{I_{y}%
=B_{0}S_{y}I_{y},}\cr\displaystyle{I_{z}=B_{0}(S_{0}I_{x}+S_{z}I_{z}%
),}\label{integal equation system}%
\end{cases}
\end{equation}
where the notations
%16
\[
\displaystyle{S_x=\int \frac{d^2k k_x^2 e^{-2 k
h}}{\Delta_e(\omega,k)}}\,, \displaystyle{S_y=\int \frac{d^2k k_y^2
e^{-2 k h}}{\Delta_e(\omega,k)}},
\]
\begin{equation}
\displaystyle{S_z=\int \frac{d^2k k^2 e^{-2 k
h}}{\Delta_e(\omega,k)}},\, \displaystyle{S_0=\int \frac{d^2k k_x k
e^{-2 k h}}{\Delta_e(\omega,k)}}\,.\label{Sj}
\end{equation}
are used for the calculable integrals. Note that the parameter
${S_{0}=0}$, if the electron drift is absent ($\mathbf{v}_{0}=0$).

The zero value of the determinant of the system of equations
(\ref{integal equation system}) is a condition for the nontrivial
solutions of the integral equation (\ref{integal equation}) to
exist. This condition, which determines the frequency $\omega$, will
be referred to as a dispersion equation. If the electron velocity is
directed along the $OX$ axis, the dispersion equation is
%17
\begin{equation}
\left[  (1-B_{0}S_{x})(1-B_{0}S_{z})-B_{0}^{2}S_{0}^{2}\right]  \left[
1-B_{0}S_{y}\right] =0.\label{main dispersion equation}%
\end{equation}
The dispersion equation (\ref{main dispersion equation}) describes all
possible collective oscillations of electrons and the dipole. If
electrons are in the equilibrium state ($\mathbf{v}_{0}=0$ and
$S_{0}=0$), the dispersion equation can be decomposed into three
equations: $(1-B_{0}S_{x})=0$,\thinspace
$(1-B_{0}S_{y})=0$,\thinspace and $(1-B_{0}S_{z})=0$. The solutions
of each of them correspond to different orientations of the induced
dipole. It is evident that the solutions of equations (the
frequencies of collective dipole and
electron oscillations), which correspond to the $x$- and $y$%
-orientations of the dipole, are identical, because these two directions are physically
equivalent in the absence of charge carrier drift. The frequency of collective
oscillations for the $z$-orientation of the dipole is, generally speaking, different
from those corresponding to the $x$- and $y$-orientations.

In the general case, Eq.~(\ref{main dispersion equation}) is factorized into
two separate equations. One of them is the equation
\[
{(1-B_{0}S_{x})(1-B_{0}S_{z})=B_{0}^{2}S_{0}^{2}}\,.
\]
One can easily verify that, for the frequencies, which are the solutions of
this equation, the electric field of electrons has a symmetry, at which the
induced dipole of a nanoparticle lies in the $x-z$ plane, i.e. it corresponds to
a mixed $x-z$ orientation. The roots of this equation are two frequency
branches. The branch, which is associated at ${v_{0}=0}$ with the
$x$-orientation of the dipole (i.e. which is determined by the equation
${B_{0}S_{x}=1}$), will be called the $x$-branch. Accordingly, the other
branch will be called the $z$-branch. As it was in the equilibrium case
(${v_{0}=0}$), the solutions of the equation
\[
{B_{0}S_{y}=1}%
\]
correspond to the orientation of the dipole in parallel to the $OY$ axis for any
$v_{0}$-value; therefore, the corresponding frequency branch will be referred
to as the $y$-branch.

The solutions of the dispersion equation (\ref{main dispersion equation}) can be
presented in the form
%18
\begin{equation}
\Omega_{j}=\Omega_{0}-i\Gamma_{0}+\frac{\Lambda}{\Omega_{0}}R_{j}(V_{0}%
,\Omega_{j},\Gamma_{p}),\label{frequency}%
\end{equation}
where the subscript $j=x,y,z$ denotes the $x$-, $y$-, and $z$-branches,
respectively. We also used the notations
\[
\label{dimensionless}
\displaystyle{\Omega_j=\frac{\omega_j}{\omega_{pl}}},\quad
\displaystyle{\omega_{pl} = \sqrt{\frac{2 \pi e^2 n_0}{\kappa m
h}}},\quad V_0=\frac{v_0}{\omega_{pl} h}\,,
\]
\[
\Omega_0=\frac{\omega_0}{\omega_{pl}},\quad
 \displaystyle{\Gamma_0=\frac{\gamma_0}{\omega_{pl}}},\quad
\displaystyle{\Gamma_p=\frac{1}{2\gamma_p\omega_{pl}}},
\]
\[
\displaystyle{\Lambda=\frac{e^2 e^{*^2} {n}_0
|\langle1|x|0\rangle|^2} {\kappa^2 m h^4 \hbar \omega_{pl}^4 }}\,.
\]
The parameter $\Lambda$ is responsible for the coupling between the
2DEG and the dipole; and ${R_{j}(V_{0},\Omega_{j},\Gamma_{p})}$ is a
certain complex-valued function depending on the electron drift
velocity, the frequency of dipole oscillations, and the plasmon
damping. For the $x$- and $z$-branches, the expressions for the
functions $R_{x,z}$ are
%19
\begin{equation}
\displaystyle{R_{x,z}=\!\frac{1}{2}\!\left[S_x\!+\!S_z
\!\pm(S_x\!-\!S_z)\sqrt{1+\!\left(\frac{2S_0}{S_x-S_z}\right)^2}
\right],\!}\label{dimensionless}
\end{equation}
where the plus sign corresponds to the $x$-branch, and the
minus sign to the $z$-branch. It is easy to obtain a simpler
expression for the $y$-branch,
\[
{R_{y}=S_{y}}\,.
\]
The real part of this function, ${R_{j}^{\prime}}$, describes a shift of
the oscillation frequency of the system with respect to the eigenfrequency of the dipole
owing to the interaction between the dipole and electrons. The
corresponding imaginary part, ${R_{j}^{\prime\prime}}$, is responsible for
the additional damping (or growing) of oscillations in the hybrid system.

From the analysis of integrals (\ref{Sj}), it follows that, in the
extremely high-frequency case (\ref{crit}), the functions ${R_{j}}$
practically do not depend on the parameter $\Gamma_{p}$. (In the hydrodynamic
approximation, which is used in this work, the plasmon damping is exclusively
governed by the one-particle relaxation of electrons. The Landau damping for
plasmons, which arises, if the electron kinetics is considered \cite{Landau}, is
assumed small.) In the absence of electron drift, it can be shown analytically
that, if ${\Gamma_{p}\rightarrow0}$,
%20
\[
 \displaystyle{R_{x, y}}=\pi \mathcal{P}\!\!\int{dk\frac{k^3e^{-2k}}{\Omega_{x,y}^2-k}
 -i\pi^2\Omega_{x,y}^6e^{-2\Omega_{x,y}^2},}
\]
\begin{equation}
 \displaystyle{R_z=2 R_{x, y}}.
\end{equation}
The nonzero value of ${R_{j}^{\prime\prime}}$ and, respectively,
the \textit{oscillation damping} (analogous to the Landau damping
\cite{Landau}) in the hybrid system originate from the collective
interaction. Really, since the spectrum of plasmons
(Eq.~(\ref{plasma})) is continuous, for any given frequency
$\omega_{0},$ there can always be found \textquotedblleft
resonance\textquotedblright\ plasmons, the charge waves of which are
proportional to $\exp[i(\mathbf{k_{r}r}-\omega t)]$, where the wave
vector
$\mathbf{k_{r}}$ satisfies the condition $\omega^{\pm}(\mathbf{k_{r}}%
)=\omega_{0}$. At the same time, the electric field of the dipole is a sum of
Fourier components $\mathbf{E}_{d,\mathbf{k}}\exp[i\mathbf{k}\mathbf{r}%
-i\omega_{0}t]$, with ${\mathbf{k}=\mathbf{k}_{r}}$ inclusive. The dipole
field waves with the wave vectors $\mathbf{k}_{r}$ and the \textquotedblleft
resonance\textquotedblright\ plasmons are cophased, and they propagate with
the same velocity. This means that the corresponding dipole field acts
permanently (without variation in time) on \textquotedblleft
resonance\textquotedblright\ plasmon charges. As a result, the work of the dipole
field over the charges is nonzero, and the dipole loses its energy. Under
nonequilibrium conditions, the dipole can acquire the energy from the electron
subsystem. A necessary condition for that is
%21
\begin{equation}
\displaystyle{R_{j}^{\prime\prime}(V_{0},\Omega_{j},\Gamma_{p})>0\,.}%
\label{inst-cond-1}%
\end{equation}
The sufficient condition for the hybrid system to be unstable in whole is more
rigorous:
%22
\begin{equation}
\displaystyle{\frac{\Lambda}{\Omega_{0}}R_{j}^{\prime\prime}(V_{0},\Omega
_{j},\Gamma_{p})>\Gamma_{0}\,,}\label{inst-cond-2}%
\end{equation}
If the inverse inequality is obeyed, either the system is stable or the
oscillations in the system attenuate.

%Fig. 2
\begin{figure*}% figure* for wide figure, [h] [!] to change the placement
\includegraphics[width=8.3cm]{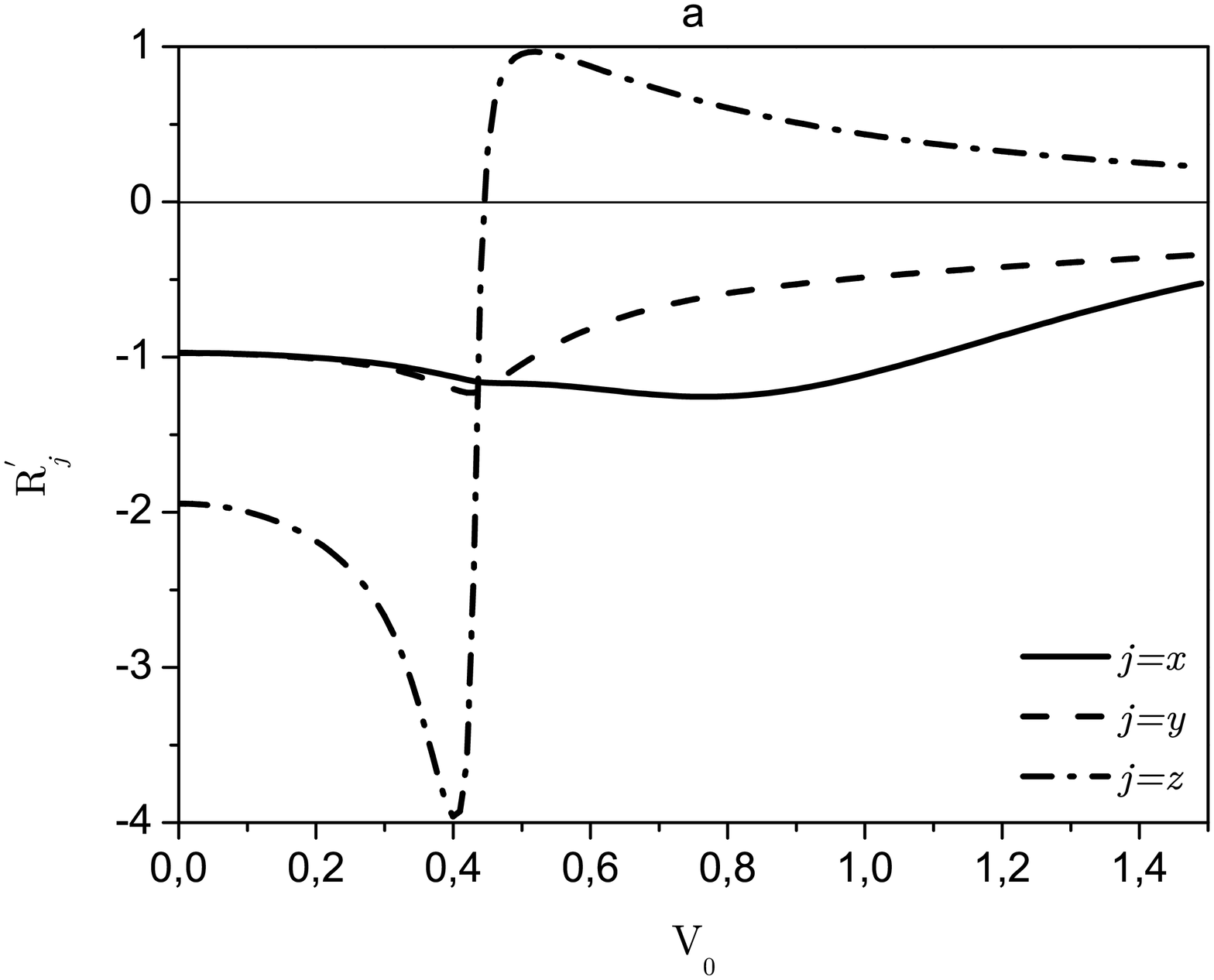}\hspace{0.5cm}\includegraphics[width=8.3cm]{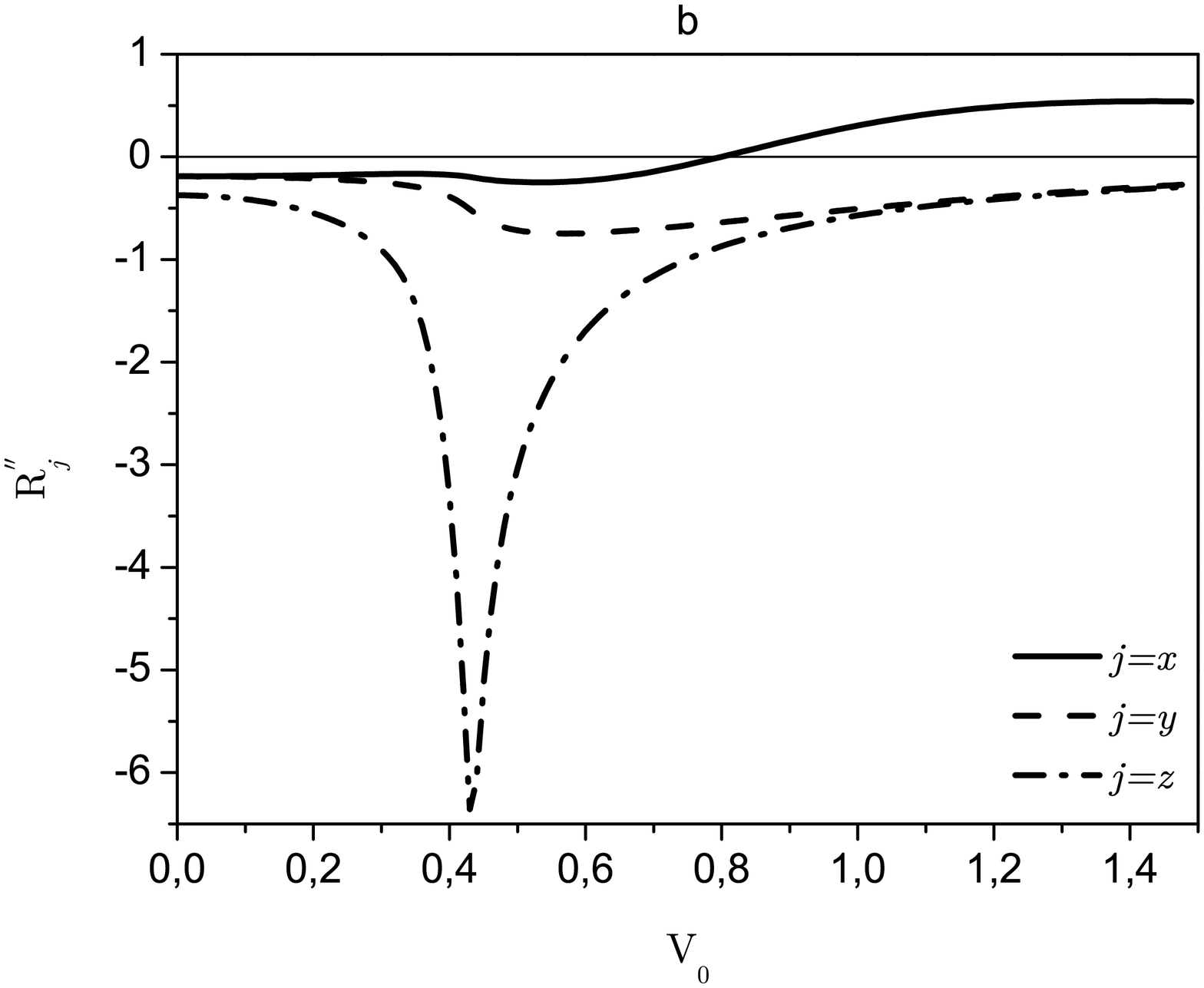}
\caption{Dependences of the real ($a$) and imaginary ($b$) parts of
the function $R_{j}$ on the electron drift velocity at a fixed
frequency of dipole oscillations (${\Omega_{0}\approx0.58}$). The
other parameters are given in the text   }\label{Fig-2}
\end{figure*}

\section{Interaction between a Shallow Donor and Drifting Two-Dimensional
Electrons}

In this section, we apply the results obtained above to a specific hybrid
system consisting of a shallow impurity center and a heterostructure with a
quantum well.

It is well known that one-particle Coulomb impurities in semiconductors are
characterized by low binding energies, and the allowed photo-induced dipole
transitions between impurity states correspond to the THz spectral range. Such
impurities can be regarded as hydrogen-like atoms, the energy spectrum and the
wave functions of which can be calculated in the effective mass approximation
(see, e.g., work \cite{Mitin}). The energy difference between the basic, $1S$,
and excited, $2P$, states is evaluated as ${E_{S-P}={3e^{2}}/{8\kappa a_{B}}}%
$, where ${a_{B}=\kappa\hbar^{2}/me^{2}}$ is the radius in the ground state.
For example, for GaAs with ${m=0.067m_{0}}${ and }${\,\kappa=12.9}$ ($m_{0}$
is the free electron mass), we obtain ${E_{S-P}\approx4.12~}$meV and
${a_{B}\approx10}$\textrm{~nm}. This energy difference corresponds to the
frequency ${\omega_{0}\approx6.2\times10^{12}}$~s$^{-1}$ ($\approx0.99$~THz).
The phototransition ${S\leftrightarrow P}$ is an allowed electric dipole
transition with the transition matrix element ${\langle1|x|0\rangle}%
\approx{0.52}\,{a_{B}^{\ast}}$. The cited parameters allow the polarizability
of a Coulomb impurity to be calculated using relation (\ref{beta0}).

To achieve conditions needed for the instability excitation, a
substance for the heterostructure must be selected, which would be
characterized by high electron velocities. Consider a quantum well
on the basis of InAs with GaAs-barriers. It is known [24] that the
effective electron mass in InAs is small, ${m\approx0.023m_{0}}$,
and the electron mobility is high even
at room temperature, ${\mu\approx8\times10^{4}}~\mathrm{cm}^{\mathrm{2}%
}/(\mathrm{V\times s})$. It gives rise to very high drift velocities
of electrons, up to ${v_{0}\approx6\times10^{7}}$~$\mathrm{cm/s}$
\cite{Kuchar,Krotkus}. In quantum wells on the basis of InGaAs,
drift velocities of the same order are observed. The difference
between the dielectric constants of the quantum well and the barrier
can be neglected \cite{Kukhtaruk_UJP}. For numerical calculations,
let us choose such physical parameters that criteria (\ref{crit})
are satisfied. In particular, let us fix the electron concentration
${n_{0}=10^{11}}~\mathrm{cm}^{-2}$ and the distance from the 2DEG to
the donor ${h=4\times10^{-6}}$~$\mathrm{cm}$. The corresponding
characteristic parameters, which were introduced by relations
(\ref{dimensionless}), are
${\omega_{pl}\approx1.07\times10^{13}}$~s$^{-1}$,
${\Lambda\approx0.0013}$, and ${\Gamma_{p}\approx0.03}$. The drift
velocity of charge carriers is normalized by the quantity
${\omega_{pl}h\approx 4.28\times10^{7}}$~$\mathrm{cm/s}$.

%Fig. 3
\begin{figure*}% figure* for wide figure, [h] [!] to change the placement
\includegraphics[width=8.3cm]{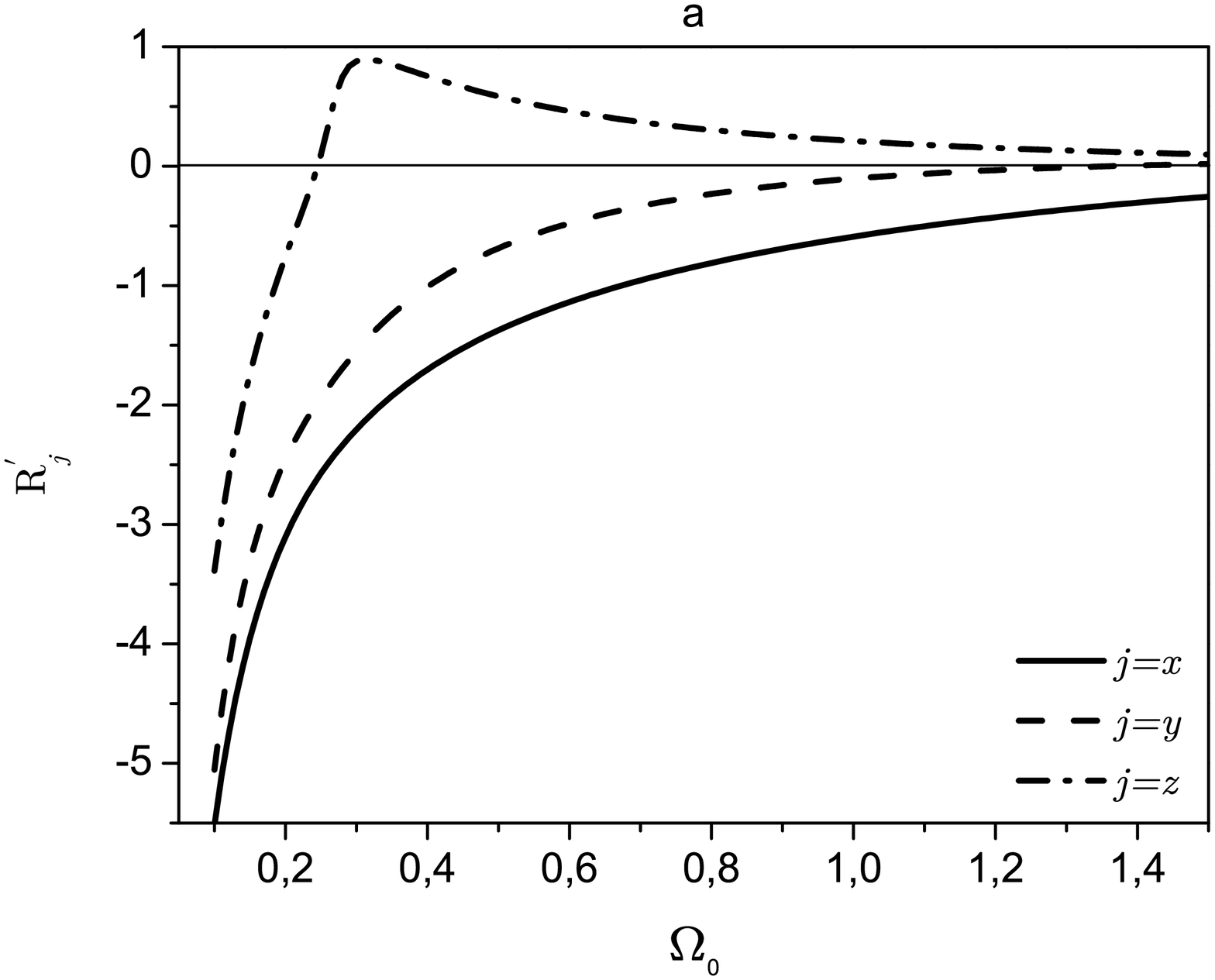}\hspace{0.5cm}\includegraphics[width=8.3cm]{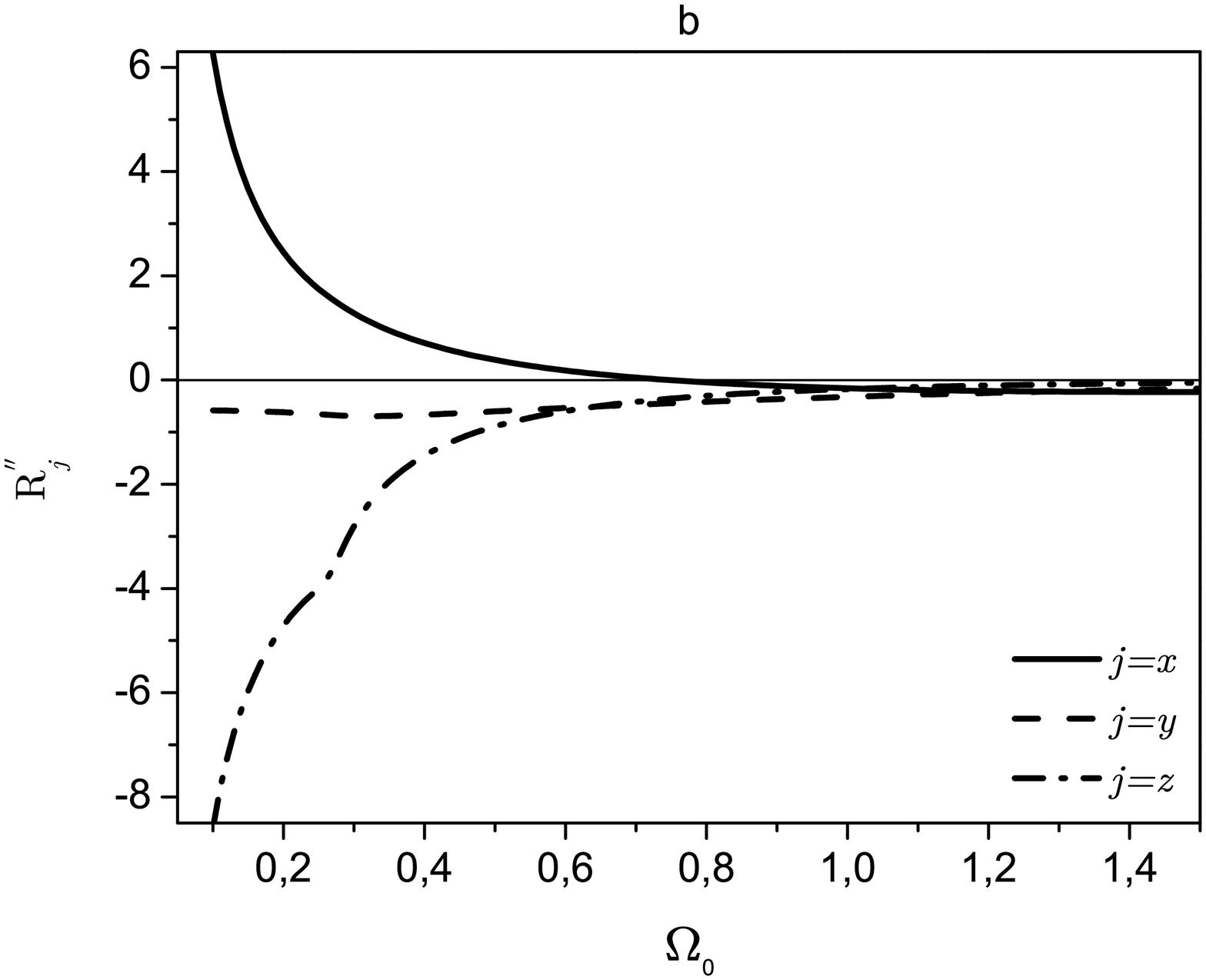}
\vskip-3mm\caption{Dependences of the real ($a$) and imaginary ($b$)
parts of the function $R_{j}$ on the frequency $\Omega_{0}$ at the
fixed electron drift velocity ${V_{0}\approx0.94}$. The other
parameters are given in the text   }\label{Fig-3}
\end{figure*}

In Figs.~2 and 3, the dependences of the real, $R_{j}^{\prime}$, and
imaginary, $R_{j}^{\prime\prime}$, parts of the function $R_{j}$ on the normalized
velocity $V_{0}$ and frequency $\Omega_{0}$, respectively, are depicted.
Figures~2,$a$ and 3,$a$ bring us to a conclusion that, for the selected fixed
parameters, the frequency of oscillations in the system is a little (in
comparison with $\omega_{0}$) shifted toward lower values in the frequency
$x$- and $y$-branches, and to both sides in the $z$-branch. Figure~2 also
demonstrates that $R_{x}^{\prime\prime}$ changes its sign at ${V_{0}%
\approx0.8}$, which corresponds to the drift velocity ${v_{0}%
\approx3.4\times10^{7}}$~$\mathrm{cm/s}$. If condition (\ref{inst-cond-2}) is
also obeyed at that, an instability must emerge, and oscillations in the
system start to grow. The figure also illustrates that the $x$- and
$y$-branches merge together at low enough drift velocities, which is
associated with the equivalence of those directions at ${V_{0}\rightarrow0}$.
Figure~3 testifies that the lower the frequency of dipole oscillations, the
larger is the instability increment. As to the $y$- and $z$-branches,
${R_{y}^{\prime\prime}<0}$ and ${R_{z}^{\prime\prime}<0}$ in the considered
ranges of frequencies and drift velocities; therefore, the corresponding
oscillations in the system attenuate here.

As was shown in works \cite{Burghoon_1994, Kalkman_1996,
Allen_2005}, the inverse lifetime of donor electrons at shallow
donors in GaAs can be of the order of $10^{7}~\mathrm{s}^{-1}$. For
${V_{0}\approx0.94}$ and ${\Omega _{0}\approx0.58}$, we obtain
${\frac{\Lambda}{\Omega_{0}}R_{x}^{\prime\prime
}\omega_{pl}\approx5.28\times10^{9}}~\mathrm{s}^{-1}$. In this case,
criterion (\ref{inst-cond-2}) is satisfied well, and the indicated
instability effects must be observed for the hybrid system
concerned. Note that the increase in the electron concentration or
the distance $h$ leads to a reduction of the instability increment,
because ${\Lambda\propto n_{0}^{-1}}$ and ${\Lambda\propto h^{-2}}$.
We also verified that  all effects discussed depend weakly on the
parameter~$\Gamma_{p}$.\looseness=1

Therefore, using an InAs quantum well and a shallow hydrogen-like donor in the
GaAs barrier as an example, we showed that an instability can take place in
the $x$-branch of collective oscillations in the system. Oscillations
corresponding to the $y$ and $z$ frequency branches attenuate.

\section{Charge Waves}

In the previous sections, the dispersion equation for collective oscillations
of the dipole and the 2DEG was derived and analyzed. Its solutions are
the characteristic frequencies (eigenvalues) of oscillations. In this
section, we are going to construct eigenfunctions, which correspond to those
eigenvalues and are the solutions of the integral equation
(\ref{integal equation}). The analysis of these solutions will allow one to
understand the behavior of the electron subsystem at collective oscillations.

Let us recall that the integral equation (\ref{integal equation}) for the charge
carrier concentration $n_{k}$ must be specified for the corresponding type of
collective oscillations, after the solutions of the system of algebraic
equations (\ref{integal equation system}) for $I_{x}$, $\,I_{y}$, and
$\,I_{z}$ (they are functionals of $n_{k}$) have been found. For two mixed
$x-z$ orientations of the dipole, the quantity $I_{y}$ is nullified. Then,
using the first and third equations of this system, it is easy to obtain
\[
\displaystyle{I_{z}=K_{x}I_{x}=K_{z}^{-1}I_{x},}%
\]
where the notations ${K_{x}=B_{0}S_{0}/(1-B_{0}S_{z})}$ and ${K_{z}=B_{0}%
S_{0}/(1-B_{0}S_{x})}$ are introduced (in these notations, the
dispersion equation for the $x$- and $z$-branches reads
${K_{x}K_{z}=1}$). Then, the solution of the linear integral
equation (\ref{integal equation}), which corresponds to the
$x$-branch of frequencies, is
%23
\begin{equation}
n_{k}^{(x)}=CB_{0}\frac{k\left(  k_{x}+K_{x}k\right)  }{\Delta_{e}(\omega
_{x},k)}e^{-kh}\,,\label{rok}%
\end{equation}
where the superscript in parentheses in the notation $n_{k}^{(x)}$ means that
the frequency entering into the quantities $B_{0}$, $K_{x}$, and $\Delta_{e}$
belongs to the $x$-branch. Generally speaking, the constant $C$ is an
arbitrary complex value characterized by an amplitude and a phase. The latter
is insignificant, because it can always be zeroed by shifting the start of
time counting. Therefore, $C$ is assumed real-valued below.

Substituting Eq.~(\ref{rok}) into the formula of the Fourier transformation, we obtain
the space-time distribution of a concentration perturbation,
%24
\begin{equation}
n^{(x)}(\mathbf{r},t)=CB_{0}\int d^{2}k\frac{k\left(  k_{x}+K_{x}k\right)
}{\Delta_{e}(\omega_{x},k)}e^{-kh+i\mathbf{kr}-i\omega_{x}t}.\label{nrok}%
\end{equation}
It is the real part of expression (\ref{nrok}) that has a physical meaning.
Let
\[
{J_x (\mathbf{r}) =C B_0 \int d^2k\frac{k k_x}
{\Delta_e(\omega_j,k)} e^{-k h + i \mathbf{k r}} }\,,
\]
\[
{J_z (\mathbf{r}) =C B_0 \int d^2k\frac{k^2} {\Delta_e(\omega_j,k)}
e^{-k h + i \mathbf{k r}}}
\]
and ${\tau_{j}=\omega_{j}^{\prime}t}$, where $j=x$ for the
$x$-branch. Extracting the real part of the concentration perturbation
(\ref{nrok}), we obtain
%25
\[
 \displaystyle{n^{\prime \: (x)}(\mathbf{r},\tau_x)= }
\]
\[
=\displaystyle{e^{\frac{\omega_x^{\prime\prime}}{\omega_x^{\prime}}\tau_x}
\biggl[(J_x^{\prime}(\mathbf{r})+K_x^{\prime} J_z^{\prime}
(\mathbf{r})-K_x^{\prime\prime} J_z^{\prime\prime}(\mathbf{r}))
\cos{\tau_x}+}
\]
\begin{equation}
 \displaystyle{+(J_x^{\prime\prime}(\mathbf{r})
+K_x^{\prime} J_z^{\prime\prime}(\mathbf{r}) +K_x^{\prime\prime}
J_z^{\prime}(\mathbf{r})) \sin{\tau_x}\biggr]}.\label{nrokRe}
\end{equation}

%Fig. 4
\begin{figure}% figure* for wide figure, [h] [!] to change the placement
\includegraphics[width=\column]{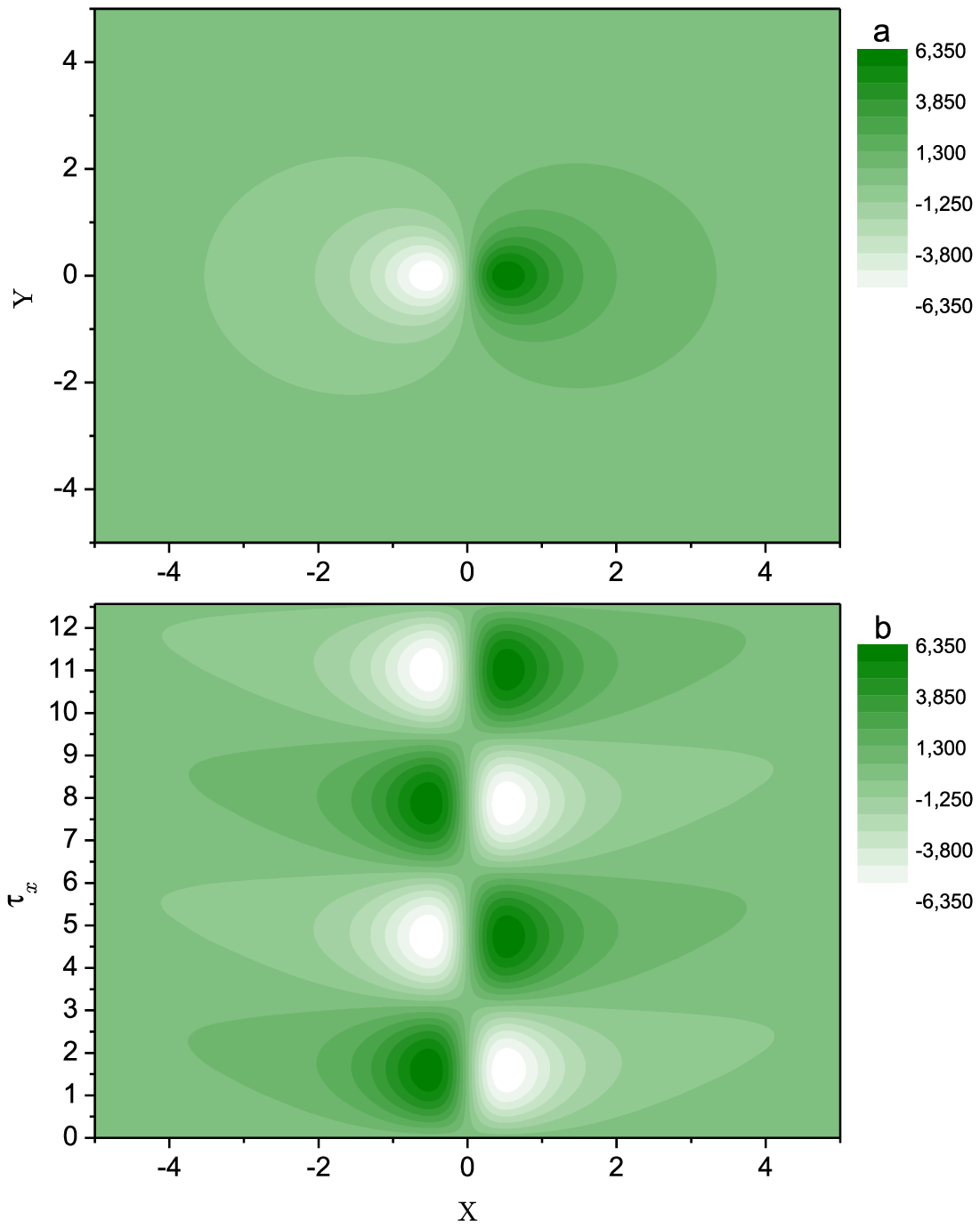}
\vskip-3mm\caption{Contour plots of the space-time dependences of
electron concentration perturbations at $V_{0}=0$, which correspond
to the $x$-branch of frequencies: ($a$) the spatial dependence at
$\tau_{x}=\frac{3\pi}{2}$, ($b$) the space-time dependence at $Y=0$
}\label{Fig-4}
\end{figure}

\noindent Solutions (\ref{nrokRe}) have the following properties.
The function $n^{\prime(x)}(\mathbf{r},\tau_{x})$, as well as
$n^{(x)}(\mathbf{r},t)$, is even with respect to $y$.\ Since
${\frac{\omega_{x}^{\prime\prime}}{\omega _{x}^{\prime}}\ll1}$, the
amplitude of this function changes slightly within several periods.
At ${V_{0}=0}$, $n^{\prime(x)}(\mathbf{r},\tau_{x})$ is an
odd function of the coordinate $x$. In general, $n^{\prime(x)}(\mathbf{r}%
,\tau_{x})$ describes the behavior of charge waves in the two-dimensional
space and in time.

%Fig. 5
\begin{figure}% figure* for wide figure, [h] [!] to change the placement
\includegraphics[width=\column]{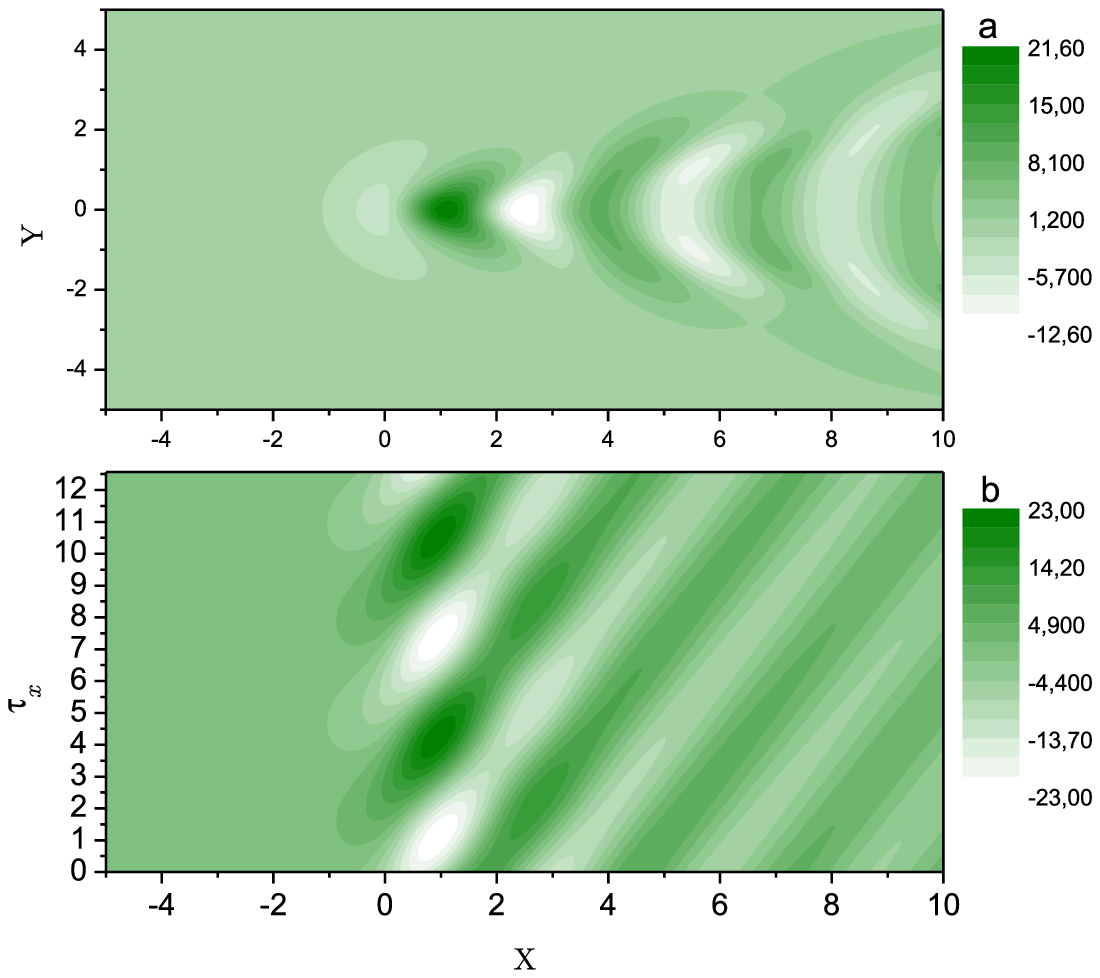}
\vskip-3mm\caption{The same as in Fig.~4, but at
${V_{0}\approx0.94}$ }\label{Fig-5}
\end{figure}

In Fig.~4, the contour plots of the space-time dependences of
electron concentration perturbations corresponding to the
$x$-branch (in this case, the dipole is oriented along the $OX$
axis) in the absence of electron drift ($V_{0}=0$) are exhibited. In
particular, in Fig.~4,$a$, the spatial distribution of the concentration
at the time moment ${\tau_{x}={3\pi}/{2}}$ is shown. It has a
minimum and a maximum to the left and to the right, respectively,
from the coordinate origin, where the dipole is located. This
distribution arose owing to the corresponding arrangement of dipole
charges at that moment. The time-space illustration (Fig.~4,$b$)
demonstrates that those maxima and minima alternate in time with a
period of $\pi$. The spatial scale of perturbations has an order of
several $h$.

The contour plots of the space-time dependences of the electron
concentration in the case where electrons drift along the $OX$ axis
(${V_{0}\approx0.94}$), are shown in Fig.~5. In the general case, if electrons
drift, the dipole becomes oriented in the plane $y=0$ (see the next section).
One can see that the concentration minima and maxima, which are exhibited in
Fig.~4, are \textquotedblleft blown\textquotedblright\ by the electron stream,
with the perturbation boundaries moving faster than its center. The
nonzero phase velocity of plasmon waves, which is
perpendicular to the electron drift, results in a smearing of perturbations
located far from the dipole.

A similar analysis of the time-space distributions of electrons can be
carried out for the $z$-branch of the frequency dispersion. For this case, instead
of Eq.~(\ref{nrokRe}), we obtain
%26
\[
\displaystyle{n^{\prime\: (z)}(\mathbf{r},\tau_z)=}
\]
\[
\displaystyle{=e^{\frac{\omega_z^{\prime\prime}}{\omega_z^{\prime}}\tau_z}\biggl[(K_z^{\prime}
J_x^{\prime}(\mathbf{r})- K_z^{\prime\prime}
J_z^{\prime\prime}(\mathbf{r})+ J_z^{\prime}(\mathbf{r}))
\cos{\tau_z}+}
\]
\begin{equation}
\displaystyle{+(K_z^{\prime} J_x^{\prime\prime}(\mathbf{r})+
K_z^{\prime\prime} J_z^{\prime}(\mathbf{r})+
J_z^{\prime\prime}(\mathbf{r}))
\sin{\tau_z}\biggr]}.\label{nrokRe_z}
\end{equation}
The properties of this function are a little similar to those of the
function ${n^{\prime(x)}(\mathbf{r},\tau_{x})}$. In particular, this
function is also
even with respect to $y$. However, in contrast to ${n^{\prime(x)}%
(\mathbf{r},\tau_{x})}$, ${n^{\prime(z)}(\mathbf{r},\tau_{z})}$ is
an even function of the coordinate $x$ at $V_{0}=0$.

%Fig. 6
\begin{figure}% figure* for wide figure, [h] [!] to change the placement
\includegraphics[width=\column]{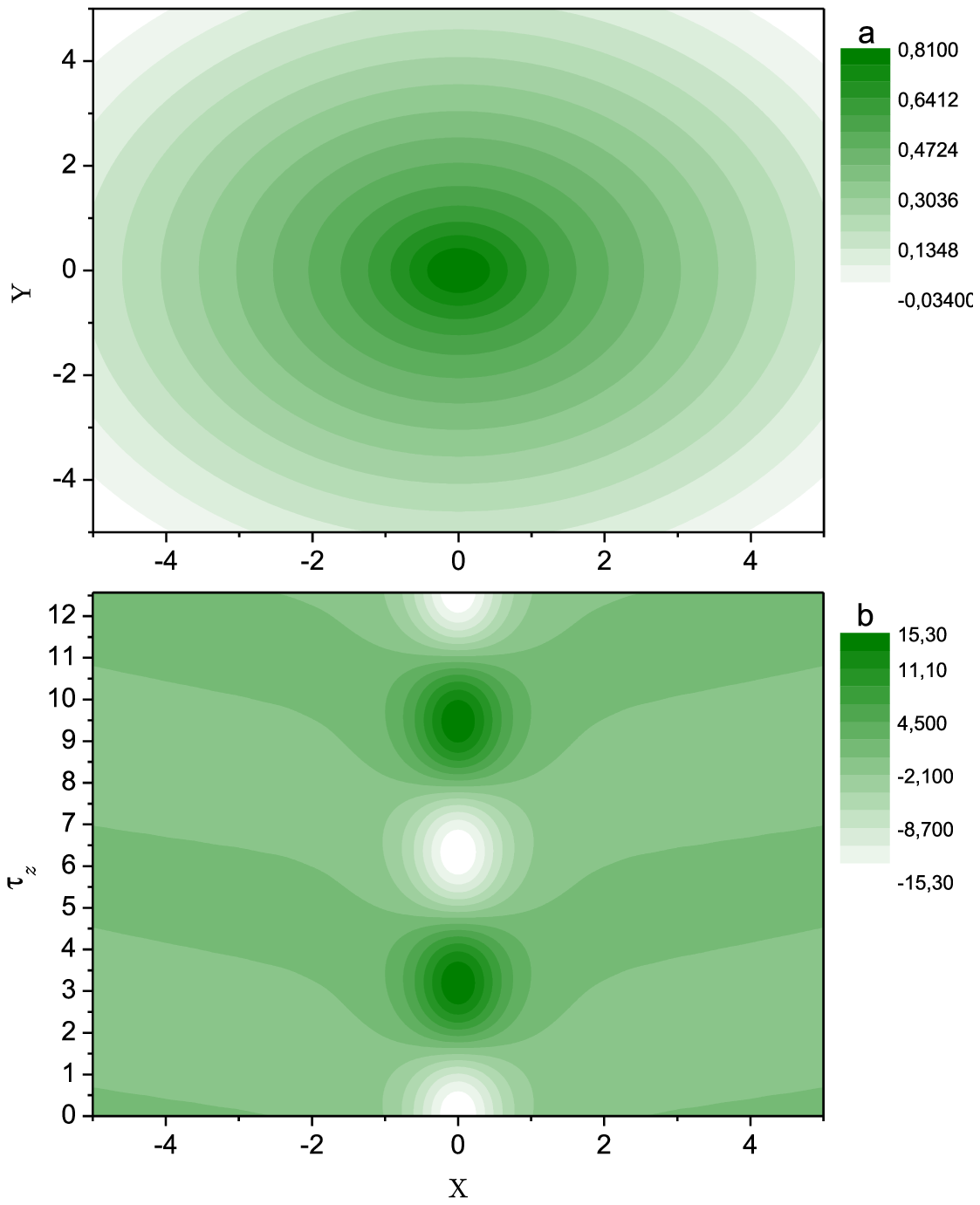}
\vskip-3mm\caption{The same as in Fig.~4, but for the
$z$-branch  }\label{Fig-6}
\end{figure}

Figure~6 illustrates the behavior of the electron concentration for the
$z$-branch of frequencies at ${V_{0}=0}$ (the dipole is oriented along the $OZ$ axis). The
spatial distribution of the electron concentration calculated for the time moment
${\tau_{x}={3\pi}/{2}}$ (see Fig.~6,$a$) has a maximum, which is located under
the dipole and is symmetric with respect to the substitutions ${x\rightarrow
-x}$ and ${y\rightarrow-y}$. Figure~6,$b$ demonstrates the behavior of
a perturbation in time, which is close to periodic. Figure~7 exhibits the same
as in Fig.~6, but provided that electrons drift with the velocity
${V_{0}\approx0.94}$. In this case, the dipole is also oriented in the plane
$y=0$, but the wave dynamics is somewhat more complicated than that obtained
above for the $x$-branch. Perturbations located near the dipole move against the
electron stream, whereas the remote perturbations move along the stream. At
${X\approx3.5}$, the character of perturbations changes, and the role of the wave
propagation becomes dominant.

Consider the solution of the integral equation (\ref{integal equation}) for the
dipole $y$-orientation. The real part of an electron concentration perturbation
looks like
%27
\begin{equation}
\displaystyle{n^{\prime(y)}(\mathbf{r},\tau_{y})=e^{\frac{\omega_{y}%
^{\prime\prime}}{\omega_{y}^{\prime}}\tau_{y}}\left[  J_{y}^{\prime
}(\mathbf{r})\cos{\tau_{y}}+J_{y}^{\prime\prime}(\mathbf{r})\sin{\tau_{y}%
}\right]  },\label{nrokRe_y}%
\end{equation}

%Fig. 7
\begin{figure}% figure* for wide figure, [h] [!] to change the placement
\includegraphics[width=\column]{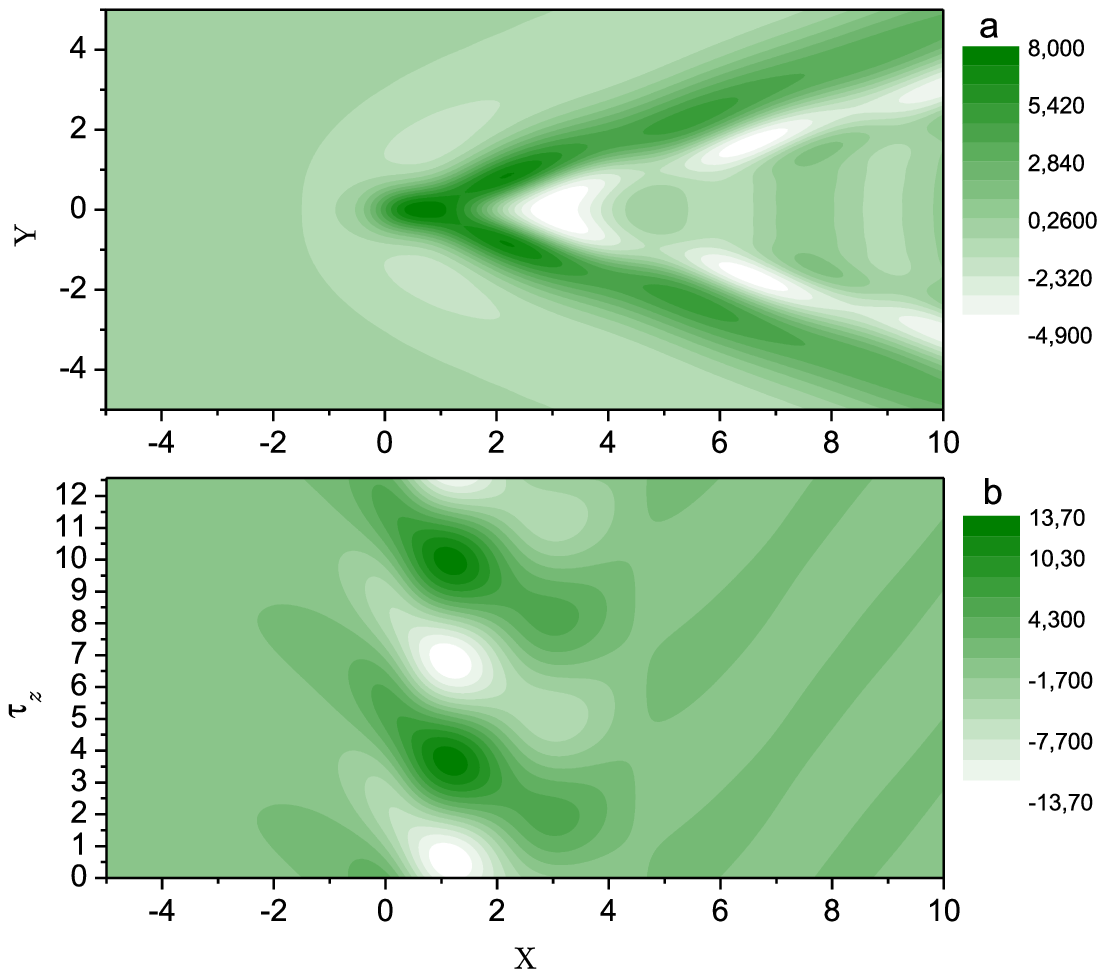}
\vskip-3mm\caption{The same as in Fig.~5, but for the
$z$-branch  }\label{Fig-7}
\end{figure}

\noindent where the notation
\[
{J_{y}(\mathbf{r})=CB_{0}\int d^{2}k\frac{kk_{y}}{\Delta_{e}(\omega_{y}%
,k)}e^{-kh+i\mathbf{kr}}}%
\]
is used. The properties of $n^{\prime(y)}(\mathbf{r},\tau_{y})$ are
also
similar with to those of $n^{\prime(x)}(\mathbf{r},\tau_{x})$. At ${V_{0}=0}%
$, the frequency branches coincide; therefore, in order to derive
$n^{\prime(y)}(\mathbf{r},\tau_{y})$ from
$n^{\prime(x)}(\mathbf{r},\tau _{x}),$ the substitution
${X\leftrightarrow Y}$ should be made. The same concerns Fig.~4 as
well. In contrast to $n^{\prime(x)}(\mathbf{r},\tau_{x})$,
$n^{\prime(y)}(\mathbf{r},\tau_{y})$ is an odd function of the
coordinate $y$ for an arbitrary $V_{0}$, which is connected with a
constant orientation of the dipole in parallel to the $OY$ axis.
Figure~8 illustrates the coordinate (panel~$a$) and coordinate-time
(panel~$b$) dependences of the electron concentration perturbation,
which corresponds to the $y$-branch at the electron
drift. At ${Y=0}$, the integrand in the integral $J_{y}%
(\mathbf{r})$ is an odd function, so that ${n^{\prime(y)}(x,y=0,\tau_{y})=0}%
$. Therefore, Fig.~8 was plotted for $y=0.5$. It can be interpreted
like Fig.~7,$b$: perturbations are blown down the electron streams;
however, in this case, perturbation minima transform into maxima and
{\it vice versa}.\looseness=1

Hence, in this section, we determined the behavior of perturbations
in the two-dimensional electron gas in the hybrid system for various
types of solutions, which takes into account whether or not the
electrons drift.\looseness=1

%Fig. 8
\begin{figure}% figure* for wide figure, [h] [!] to change the placement
\includegraphics[width=\column]{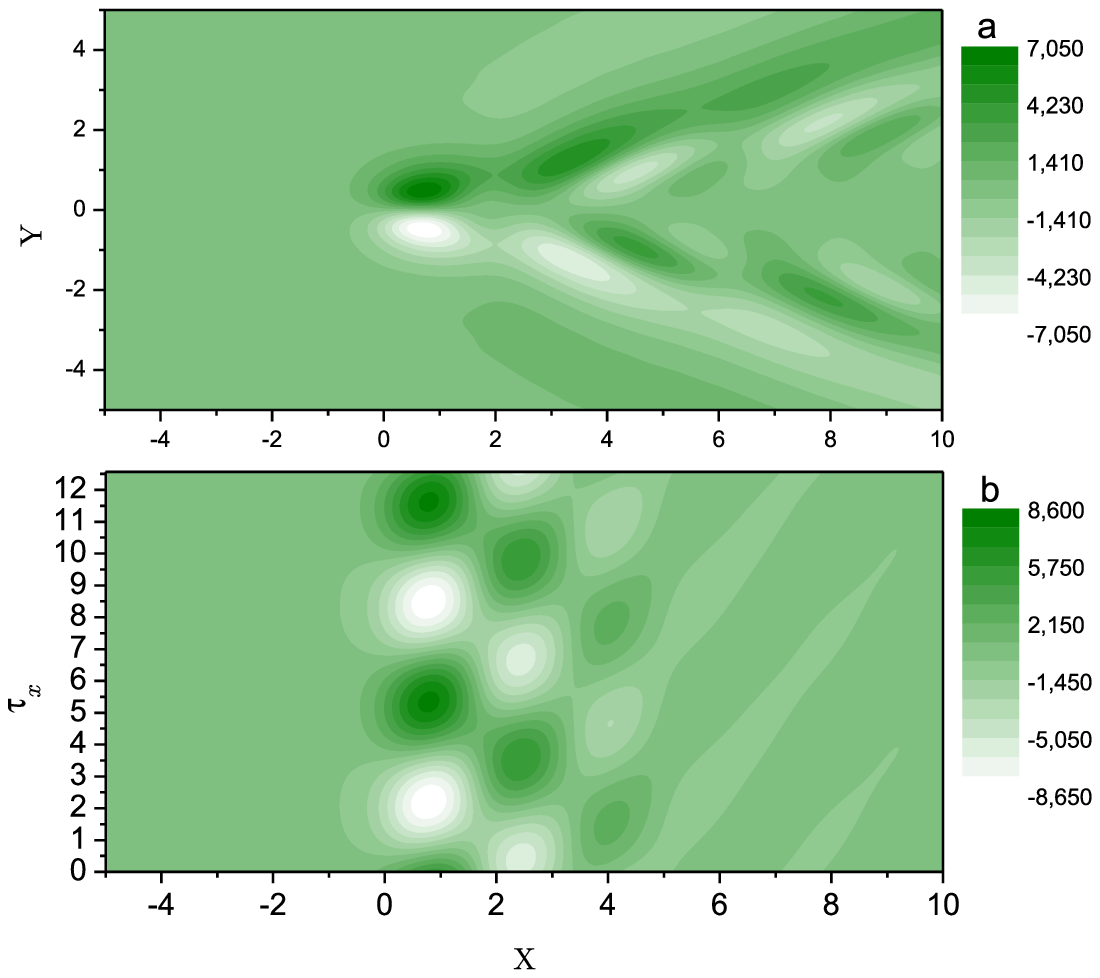}
\vskip-3mm\caption{The same as in Fig.~5, but for the
$y$-branch at $Y=0.5$  }\label{Fig-8}
\end{figure}

\section{Behavior of the Dipole Induced in a Nanoparticle}

Consider the behavior of the nanoparticle polarization during collective
electric oscillations in the hybrid system. This behavior is described by the
dynamics of the induced dipole moment (\ref{dipole}). To determine the dipole
moment, we have to calculate the electric field created by electrons at the
dipole localization point making use of definition (\ref{phi_e}), the first
formula in system (\ref{3eq}), and the Fourier components of the concentration
$n_{k}$ (the latter were analyzed for various dispersion branches in the
previous section).

First, let us consider the solutions corresponding to the dispersion
$x$-branch, provided that the components of the dipole moment are
\begin{equation}
%28
\begin{cases}
\displaystyle{d_{x}(\omega_{x})=-iC\frac{2\pi e\beta_{0}(\omega_{x})}{\kappa
}\equiv\widetilde{C}_{x}},\cr\displaystyle{d_{y}(\omega_{x})=0,}%
\cr\displaystyle{d_{z}(\omega_{x})=i\widetilde{C}_{x}K_{x}}%
,\label{dipolesys}%
\end{cases}
\end{equation}
where $\widetilde{C}_{x}$ is a new arbitrary constant. Whence, we obtain the
law for the time evolution of dipole moment components,
%29
\begin{equation}
\begin{cases}
\displaystyle{D_x=e^{\frac{\omega_x^{\prime\prime}}{\omega_x^{\prime}}\tau_x}
(\widetilde{C}_x^{\prime}\cos\tau_x+\widetilde{C}_x^{\prime\prime}\sin\tau_x),}\cr
\displaystyle{D_z=-e^{\frac{\omega_x^{\prime\prime}}{\omega_x^{\prime}}\tau_x}
\left[(\widetilde{C}_x^{\prime}K_x^{\prime\prime}
+\widetilde{C}_x^{\prime\prime}K_x^{\prime})\cos\tau_x+\right.}\cr
\displaystyle{+\left.(\widetilde{C}_x^{\prime\prime}K_x^{\prime\prime}
-\widetilde{C}_x^{\prime}K_x^{\prime})\sin\tau_x\right]}.\label{dipole_x}
\end{cases}
\end{equation}
System (\ref{dipole_x}) is a parametric equation of an ellipse: the end of
the dipole moment vector moves along an elliptic trajectory in the plane
$y=0$. It is easy to verify that ${K_{x}^{\prime}>0}$ and ${K_{x}%
^{\prime\prime}>0}$ for the $x$-branch of the frequency dispersion; hence, this
circulation is counter-clockwise. Since the dipole is in a self-consistent
field with drifting electrons, the parameters of this ellipse depend on the
drift velocity and the frequency of dipole oscillations.

Let us put the arbitrary constant ${\widetilde{C}_{x}=1}$ and, for simplicity,
let $\exp\left(  \frac{\omega_{x}^{\prime\prime}}{\omega_{x}^{\prime}}\tau
_{x}\right)  \approx{1}$, i.e. we consider the behavior of the system within several
periods. Then,
%30
\begin{equation}%
\begin{cases}
\displaystyle{D_{x}=\cos\tau_{x},}\cr\displaystyle{D_{z}=K_{x}^{\prime}%
\sin\tau_{x}-K_{x}^{\prime\prime}\cos\tau_{x}}.\label{dipole_x-z}%
\end{cases}
\end{equation}
After excluding the parameter $\tau_{x}$ from system (\ref{dipole_x-z}), we
obtain the equation for this ellipse in the form
%31
\begin{equation}
\displaystyle{aD_{x}^{2}+2bD_{x}D_{z}}+cD_{z}^{2}=1,\label{nonkanon}%
\end{equation}
where ${a=1+\left(  {K_{x}^{\prime\prime}}/{K_{x}^{\prime}}\right)  ^{2}}$,
${b={K_{x}^{\prime\prime}}/{K_{x}^{\prime^{2}}}}$, and ${c={1}/{K_{x}%
^{\prime^{2}}}}$. The rotation of the coordinate system by an angle
$\alpha_{x}$ determined by the equation
%32
\begin{equation}
\displaystyle{\tan\alpha_{x}=\frac{1-K_{x}^{\prime^{2}}-K_{x}^{\prime
\prime^{2}}}{2K_{x}^{\prime\prime}}\pm\sqrt{\left(  \frac{1-K_{x}^{\prime^{2}%
}-K_{x}^{\prime\prime^{2}}}{2K_{x}^{\prime\prime}}\right)  ^{2}+1}%
}\,,\label{tg}%
\end{equation}
brings about the canonical ellipse equation,
%33
\begin{equation}
\displaystyle{\frac{D_{x_{0}}^{2}}{A^{2}}+\frac{D_{z_{0}}^{2}}{B^{2}}%
=1,}\label{kanon}%
\end{equation}
where $D_{x_{0}}$\thinspace and\thinspace$D_{z_{0}}$ are the dipole components
in the new coordinate system, and $A$ and $B$ are the principal ellipse axes,
%34
\[
\displaystyle{A^2=\frac{1+\tg^2\alpha_x}{a+2b\tg\alpha_x+c\,\tg^2\alpha_x}},
\]
\begin{equation}
\displaystyle{B^2=\frac{1+\tg^2\alpha_x}{a\tg^2\alpha_x-2b\tg\alpha_x+c}}\,.
\end{equation}
The parameters $A$ and $B$ are used to introduce the ellipse
eccentricity,
%35
\begin{equation}
\displaystyle{\epsilon=}
\begin{cases}
\displaystyle{\sqrt{1-\frac{B^2}{A^2}}, \;\;  A>B,} \\[2mm]
\displaystyle{\sqrt{1-\frac{A^2}{B^2}},\;\;
A<B}.\label{eccentricity}
\end{cases}
\end{equation}

%Fig. 9
\begin{figure*}% figure* for wide figure, [h] [!] to change the placement
\includegraphics[width=8.3cm]{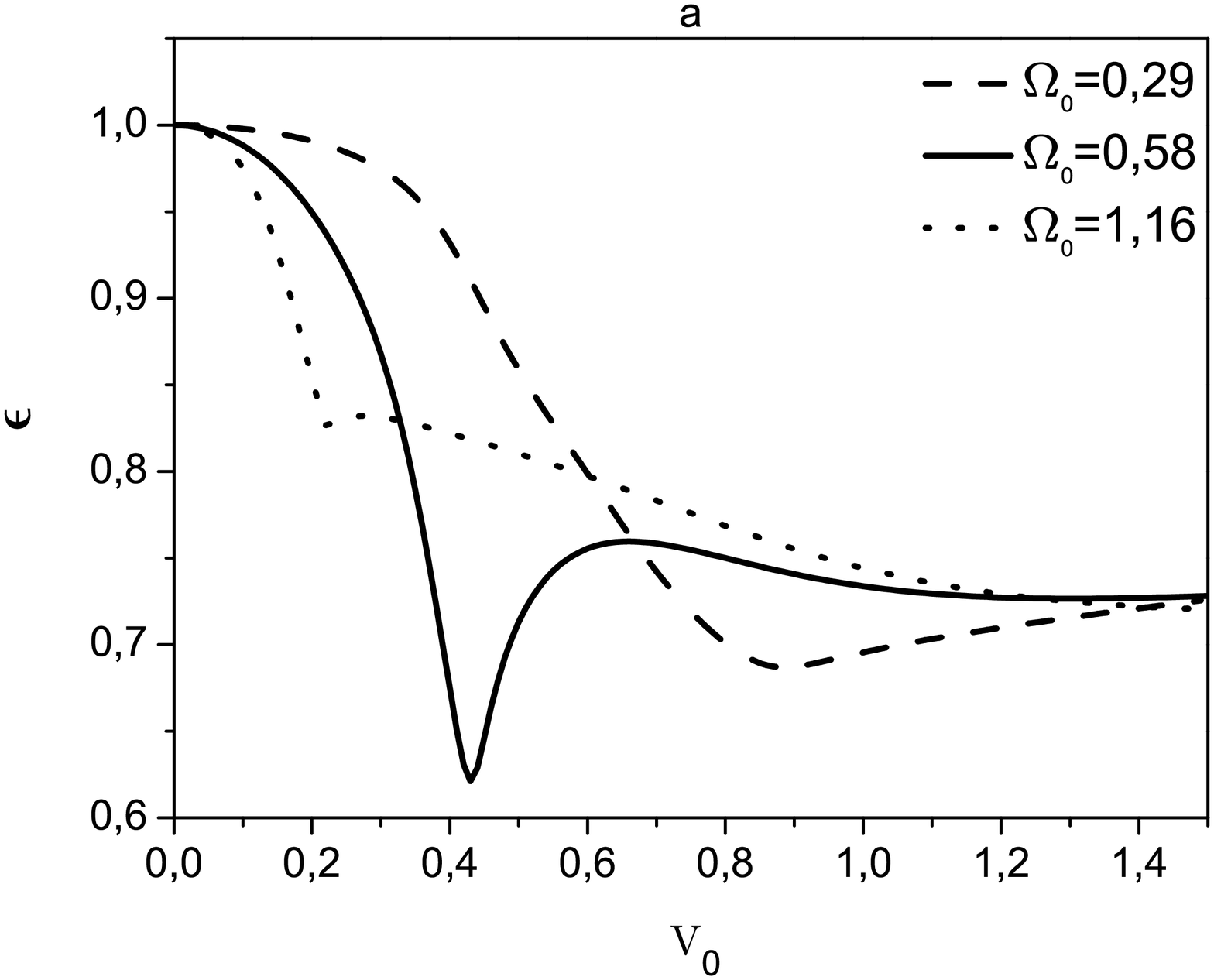}\hspace{0.5cm}\includegraphics[width=8.3cm]{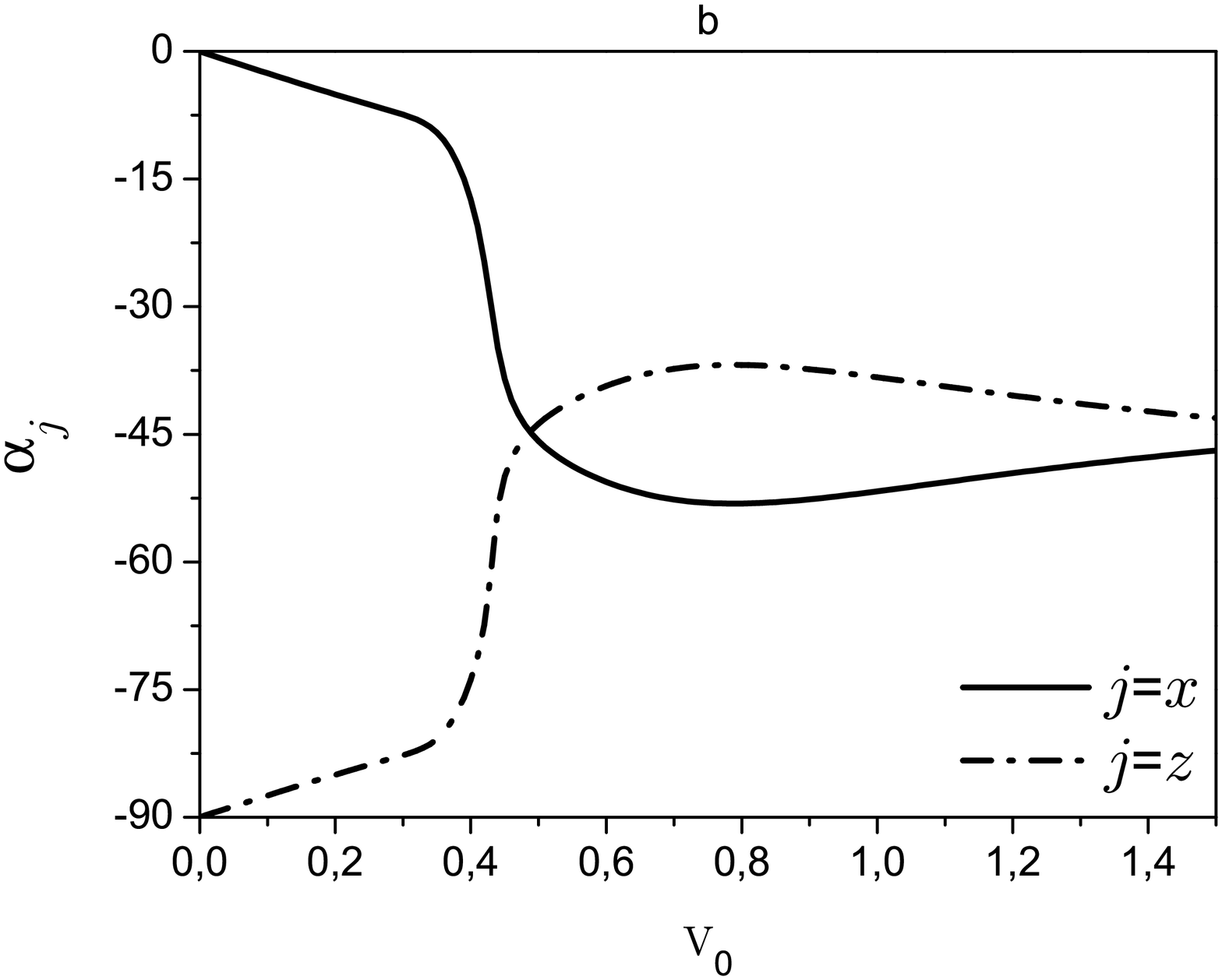}
\caption{Dependences of ($a$)~the ellipse eccentricity
$\epsilon_{j}$ and ($b$)~the rotation angle $\alpha_{j}$ (in degree
units) of the principal axes of the ellipse on the electron drift velocity
}\label{Fig-9}
\end{figure*}

The ${\epsilon}$-magnitude varies from zero to 1. The ellipse transforms
into a circle at ${\epsilon\rightarrow0}$ and into a line at ${\epsilon
\rightarrow1}$. The eccentricity of the ellipse, along which the dipole
circulates, evidently depends on the electron drift velocity and the frequency
of dipole oscillations, because the principal axes depend on $K_{x}$.

Note that the time dependence of the dipole is very simple in the coordinate
system coupled with the principal axes, namely,
%36
\begin{equation}%
\begin{cases}
\displaystyle{D_{x_{0}}=A\cos\tau_{x},}\cr\displaystyle{D_{z_{0}}=B\sin
\tau_{x}.}\label{kanon_dipole_x-z}%
\end{cases}
\end{equation}
Similar formulas can also be derived for solutions associated with
the dispersion $z$-branch. In this case, the dipole also circulates
along an ellipse, but now \textquotedblleft
clockwise\textquotedblright. The principal axes of this ellipse
depend on $K_{z}(\omega_{z})$. To obtain the canonical ellipse
equation for the $z$-branch, the substitutions
${a\leftrightarrow c}$, ${b\rightarrow-b}$, and ${K_{x}(\omega_{x}%
)\rightarrow-K_{z}(\omega_{z})}$ should be made in the equations written
above. Now, the eccentricity is described by expression (\ref{eccentricity}),
but the principal axes swap (${A\leftrightarrow B}$), and the angle
${\alpha_{x}}$ changes to ${\alpha_{z}}$. The eccentricity is the same for
both dispersion branches, because ${K_{x}(\Omega_{x})\simeq-K_{z}(\Omega_{z}%
)}$.

Concerning the $y$-orientation of the dipole, the eccentricity of the corresponding
ellipse is equal to 1 for all $V_{0}$, which is associated with the dipole
orientation strictly along the $OY$ axis for any $V_{0}$.

The dependences of eccentricity $\epsilon$ on the electron drift velocity for
three different dipole oscillation frequencies are depicted in Fig.~9,$a$. The
solid curve exposes the eccentricity that corresponds to the same frequency as
in all previous figures, where this frequency was fixed, i.e. ${\Omega
_{0}\approx0.58}$. Two other curves correspond to half as high and twice as
high dipole oscillation frequencies. The figure testifies that the
eccentricity is equal to 1, if the drift velocity equals zero, i.e. the
dipole oscillates with the frequency $\Omega_{x}$ along the $OX$ axis and with
the frequency $\Omega_{z}$ along the $OZ$ axis. Figure~9,$b$ shows that the
lines, along which the dipole oscillates, are perpendicular to each other at
${V_{0}=0}$. If $V_{0}$ increases, the eccentricity decreases, and the line
transforms into an ellipse. The ellipse, which corresponds to the $x$-branch,
rotates \textquotedblleft clockwise\textquotedblright; therefore, the absolute
value of the angle, by which its axes are rotated with respect to the $OX$
axis, grows. The other ellipse rotates \textquotedblleft
counter-clockwise\textquotedblright; therefore, the absolute value of its
orientation angle decreases. At a certain drift velocity, those ellipses coincide.

A similar situation takes place at other dipole oscillation frequencies as
well. The curves in Fig.~9 also demonstrate that the eccentricities have a
similar structure at different frequencies, and the smallest $\epsilon$-value
is realized for a curve that corresponds to ${\Omega_{0}\approx0.58}$. At this
frequency, the function $R_{j}$, the plot of which is exhibited in Fig.~2,
changes drastically in a vicinity of the drift velocity, which corresponds to
the eccentricity minimum (${V_{0}\approx0.43}$). The plots of the functions
$\alpha_{x}$ and $\alpha_{z}$ intersect each other also in a vicinity of
${V_{0}\approx0.43}$. At other frequencies, such eccentricity minima take
place at different drift velocities, and the behavior of the function $R_{j}$
in vicinities of those minima (to be exact, in the region, where the
eccentricity diminishes sharply) is also drastic.

Let us elucidate the physical reason for the nontrivial behavior
demonstrated by the induced nanoparticle dipole. For this purpose,
consider the electric field of plasmons at a certain point
$\left(x=y=0, z=h\right)$ remote from the 2DEG. The dispersion law
(\ref{plasma}) has two branches. However, for simplicity, let us
consider any of them, omitting the subscript at the frequency and
neglecting the plasmon damping. Let a plasmon with the frequency
$\omega(k)$ and the two-dimensional wave vector $\mathbf{k}=(k,0)$
be excited. Then, at the distance $h$ from the coordinate origin, it
creates a field
\begin{eqnarray}
\begin{cases}
\displaystyle{E_x=-i k |A_0| e^{-kh} e^{-i \omega t}},\cr
\displaystyle{E_z=k |A_0| e^{-kh} e^{-i\omega t }}\,, \nonumber
\end{cases}
\end{eqnarray}
where $\left\vert A_{0}\right\vert $ is an arbitrary amplitude. Only
the real part of the found field has a physical meaning. Its
components are
\begin{eqnarray}
\begin{cases}
\displaystyle{E^{\prime}_x=-k |A_0| e^{-kh} \sin{(\omega t )}},\cr
\nonumber \displaystyle{E^{\prime}_z=k |A_0| e^{-kh} \cos{(\omega t
)}}. \nonumber
\end{cases}\label{Replasmafield}
\end{eqnarray}
We see that, at a certain distance from the coordinate origin, the
field created by plasmons circulates along a circle. It is evident
that the dipole moment of a nanoparticle, which is given by formula
(\ref{dipole}), when responding to an external force, will also
circulate under the influence of the plasmon field. When collective
oscillations take place in the hybrid system, not a single but
many plasmons with different $\mathbf{k}$ are excited. The
total field has a more complicated time dependence, which results in
the dipole circulation along an ellipse. Hence, the dynamics of
the plasmon field is a key for understanding the behavior of
the nanoparticle polarizability.\looseness=1

It is important that the behavior of the induced dipole in time can be
experimentally observed owing to features of the emission
by the hybrid system. Really, having solved the problem of
an electrostatically coupled nanoparticle and the two-dimensional electron
gas, we determined the motion of charges in the hybrid system. The
motions of polarization charges in the nanoparticle and electrons
in the quantum well correspond to an electric current depending
on the coordinates and the time. The alternating current generates
radiation. Let the current density be designated as
$\mathbf{j}(x,y,z,t)$. Then, the vector potential of the radiation
field can be determined by the
formula~\cite{Landau_2}\looseness=1
%37
\begin{equation}
\displaystyle{\mathbf{A}(x,y,z,t)=\frac{1}{c_{0}}\int{\!dx_{1}dy_{1}%
dz_{1}\frac{\mathbf{j}(x_{1},y_{1},z_{1},t-\frac{R}{c_{0}})}{R}}},\label{A}%
\end{equation}

\noindent
where $c$ is the speed of light, and $R=[(x_{1}-x)^{2}+(y_{1}-y)^{2}%
+(z_{1}-z)^{2}]^{1/2}$. Using formula (\ref{A}) for the far radiation zone,
we obtain the Fourier components of the vector potential in the form
%38
\begin{equation}
\displaystyle{\mathbf{A}(x,y,z,\omega_{j})=-\frac{ik_{0}\mathbf{d}(\omega
_{j})e^{-i\omega_{j}t+ik_{0}R_{0}}}{R_{0}}}\,,\label{A_d}%
\end{equation}
where $R_{0}=[x^{2}+y^{2}+(z-h)^{2}]^{1/2}$ is the distance to
the observation point, and ${k_{0}=\omega_{j}/c_{0}}$. The frequency
$\omega_{j}$ belongs to one of three frequency branches. The dipole
moment, which determines $\mathbf{A}$, is formed by the
self-consistent system composed of the electrons and the
nanoparticle. Hence, the radiation characteristics substantially
depend on such parameters of the system as the distance between the
dipole and the quantum well, the concentration of electrons, the
frequency of dipole oscillations, and the electron drift velocity.
At finite drift velocities of electrons, the induced dipole
circulates along an ellipse, and this fact must be reflected in the
radiation polarization.\looseness=1

\section{Conclusions}

A hybrid system consisting of an isotropic nanoparticle and a heterostructure
with a quantum well has been considered. The nanoparticle is assumed to be
polarizable in an external electric field and to have characteristic resonance
frequencies in the terahertz range. In this case, there emerge collective
oscillations of the nanoparticle polarization and plasmons in the two-dimensional
electron gas of the hybrid system.

Dispersion relations for the collective oscillation frequencies were
obtained and analyzed. Possible frequency branches were determined
and classified. Additional damping of predicted oscillations was
revealed, the origin of which is similar to that of Landau
collective damping in plasma. The electron drift results in a
reduction of the additional damping. At sufficiently high  drift
velocities, owing to the energy of the electric current, there
emerges an instability in the system, and the oscillations in one of
the dispersion branches grow in time. The electric instability
increment increases, when the distance between the dipole and the
electrons diminishes and when the drift velocity increases.

The predicted effects were illustrated using the results of numerical
calculations for a shallow hydrogen-like donor in the barrier of an InAs-based
heterostructure with GaAs barriers, which were taken as an example. The
space-time dependences of concentration perturbations in the
two-dimensional electron gas in the course of collective oscillations were
analyzed. The calculated behavior was demonstrated to be substantially
different for different frequency branches, both in the absence and the presence
of drifting electrons.

The polarization oscillations of a nanoparticle were studied. It was found that
the induced dipole is characterized by a complicated dynamics at nonzero drift
velocities. In particular, in two of three branches, the dipole circulates
along elliptic trajectories depending on the electron drift parameter. It was
shown that the features in the nanoparticle polarization behavior could be
observed by measuring the emission of the hybrid system.

The practical interest to the new phenomena in hybrid systems may
consist in a capability to excite the emission by
nanoparticles by applying an electric current and the electrically
stimulated generation of THz radiation. These phenomena can also be
used for the field-controlled addressing to individual nanoparticles,
which is a key problem at the implementation of quantum calculations
\cite{Nielsen}.
\vskip3mm
 The authors express their sincere
gratitude to M.V.~Strikha for his attentive reading of the paper
and valuable remarks. The work was partially supported by the State
goal-oriented scientific and technical program \textquotedblleft
Nanotechnologies and nanomaterials\textquotedblright.

\rezume{%
ВЗАЄМОДІЯ ІЗОТРОПНОЇ НАНОЧАСТИНКИ\\ З ДРЕЙФУЮЧИМИ ЕЛЕКТРОНАМИ\\ У
КВАНТОВІЙ ЯМІ}{В.О. Кочелап, С.М. Кухтарук} {Розглянуто гібридні
системи, що складаються з наночастинки та напівпровідникової
гетероструктури з квантовою ямою. Наночастинка є такою, що
поляризується у сторонньому електричному полі. Обґрунтовано та
сформульовано модель гібридної системи. Отримано точні розв'язки
рівнянь. Знайдені частоти коливань зарядів гібридної системи та їх
додаткове загасання, що зумовлено взаємодією диполя з плазмонами.
Природа додаткового загасання подібна до загасання Ландау.
Проаналізовано поведінку в часі та просторі збурень концентрації
двовимірних електронів. Досліджено поляризаційні коливання
наночастинки. Знайдено, що при ненульових дрейфових швидкостях
наведена поляризація характеризується складною динамікою. Зокрема,
для двох із трьох гілок частотної дисперсії вектор поляризації
обертається по еліптичних траєкторіях. У випадку, коли до квантової
ями прикладене поле та тече струм, загасання змінюється на
наростання коливань гібридної системи у часі, що відповідає
електричній нестійкості гібридної системи. Нові явища в гібридних
системах можуть бути застосовані для збудження випромінювання
наночастинок струмом та для\linebreak електричної генерації
випромінювання в терагерцовій області спектра.}


\begin{thebibliography}{99}                                                                                               %

%1
\bibitem {rev-1}P.~Bakshi and K.~Kempa, Superlatt. Microstruct. \textbf{17},
363 (1995).\vskip3mm
%2
\bibitem {rev-2}S.A.~Mikhailov, Recent Res. Devel. Appl. Phys. \textbf{2}, 65 (1999).\vskip3mm
%3
\bibitem {Wilkins}B.Y.K.~Hu and J.W.~Wilkins, Phys. Rev.~B\textbf{ 41}, 10706 (1990).\vskip3mm
%4
\bibitem {Gribnikov}Z.S.~Gribnikov, N.Z.~Vagidov, and V.V.~Mitin, J.~Appl.
Phys. \textbf{88}, 6736 (2000).\vskip3mm
%5
\bibitem {f-length-1}K.~Kempa, P.~Bakshi, and E.~Gornik, Phys. Rev. B
\textbf{54}, 8231 (1996).\vskip3mm
%6
\bibitem {Dyakonov}M.~Dyakonov and M.S.~Shur, Phys. Rev. Lett. \textbf{71},
2465 (1993); Appl. Phys. Lett. \textbf{87}, 111501 (2005).\vskip3mm
%7
\bibitem {Knap-2004}W.~Knap, J.~Lusakowski, T.~Parenty, S.~Bollaert, A.~Cappy,
V.V.~Popov, and M.S.~Shur, Appl. Phys. Lett. \textbf{84}, 2331
(2004).\vskip3mm
%8
\bibitem {Knap-2005}J.~Lusakowski, W.~Knap, N.~Dyakonova, L.~Varani,
J.~Mateos, T.~Gonzalez, Y.~Roelens, S.~Bollaert, and A.~Cappy,
J.~Appl. Phys. \textbf{97}, 064307 (2005); N.~Dyakonova, A.~El
Fatimy, J.~Lusakowski, W.~Knap, M.I.~Dyakonov, M.-A.~Poisson,
E.~Morvan, S.~Bollaert, A.~Shchepetov, Y.~Roelens, Ch.~Gaquiere,
D.~Theron, and A.~Cappy, Appl. Phys. Lett. \textbf{88}, 141906
(2006).\vskip3mm
%9
\bibitem {Knap-2008}T.~Otsuji, Y.M.~Meziani, T.~Nishimura, T.~Suemitsu,
W.~Knap, E.~Sano, T.~Asano and V.V.~Popov, J.~Phys.: Condens. Matter
\textbf{20}, 384206 (2008).\vskip3mm
%10
\bibitem {Demel_1990}T.~Demel, D.~Heitman \textit{et al}., Phys. Rev. Lett.
\textbf{64}, 788 (1990).\vskip3mm
%11
\bibitem {old-QDs-1}Ch.~Sikorski and U.~Merkt, Phys. Rev. Lett. \textbf{62},
2164 (1989); B.~Meurer, D.~Heitmann, and K.~Ploog, Phys. Rev. Lett.
\textbf{68}, 1371 (1992); D.~Heitmann and J.P.~Kotthaus, Phys.
Today, \textbf{48}, 56 (1993).\vskip3mm
%12
\bibitem {old-QDs-2}S.M.~Reimann and M.~Manninen, Rev. Mod. Phys. \textbf{74},
1283 (2002).\vskip3mm
%13
\bibitem {old-QDs-3}C.P.~Garcia, S.~Kalliakos, V.~Pellegrini, A.~Pinczuk,
B.S.~Dennis, L.N.~Pfeiffer, and K.W.~West, Appl. Phys. Lett.
\textbf{88}, 113105 (2006).\vskip3mm
%14
\bibitem {Yu_2004}B.~Yu, F.~Zeng \textit{et al}., Biophys.~J. \textbf{86}, 1649 (2004).\vskip3mm
%15
\bibitem {Maistrenko_1999}V.N.~Maistrenko, S.V.~Sapernikova \textit{et al}., J.~Analyt.
Chem. \textbf{55}, 586 (2000).\vskip3mm
%16
\bibitem {Balu_1999}R.~Balu, H.~Zhang \textit{et al}., Biophys.~J. \textbf{94}, 3217 (2008).\vskip3mm
%17
\bibitem {Burghoon_1994}J.~Burghoon, T.O.~Klaassen, and W.T.~Wenchebach,
Semicond. Sci. Technol. \textbf{9}, 30 (1994).\vskip3mm
%18
\bibitem {Kalkman_1996}A.J.~Kalkman, H.P.M.~Pellemans, T.O.~Klaassen, and
W.T.~Wencheback, Int. J. Infrared Millim. Waves \textbf{17}, 569
(1996).\vskip3mm
%19
\bibitem {Allen_2005}D.G.~Allen, M.S.~Sherwin, and C.R.~Stanley, Phys. Rev.~B
\textbf{72}, 035302 (2005).\vskip3mm
%20
\bibitem {Kukhtaruk_UJP}S.M.~Kukhtaruk, Ukr. J. Phys. \textbf{55}, 8, 916 (2010).\vskip3mm
%21
\bibitem {Davydov}A.S.~Davydov, \textit{Quantum Mechanics} (Pergamon Press,
New York, 1976).\vskip3mm
%22
\bibitem {Landau}E.M.~Lifshitz and L.P.~Pitaevskii, \textit{Physical Kinetics}
(Pergamon Press, New York, 1981).\vskip3mm
%23
\bibitem {Mitin}V.V.~Mitin, V.A.~Kochelap, and M.A.~Stroscio, \emph{Quantum
Heterostructures} (Cambridge Univ. Press, New York, 1999).\vskip3mm
%24
\bibitem {Kuchar}F.~Kuchar, G.~Bauer, and H.~Hillbrand, Phys. Status Solidi A
\textbf{17}, 491 (1973).\vskip3mm
%25
\bibitem {Krotkus}A.~Krotkus and Z.~Dobrovolskis, \textit{Electrical
Conductivity of Narrow-Gap Semiconductors} (Mokslas, Vilnius, 1988)
(in Russian).\vskip3mm
%26
\bibitem {Masselink}W.T.~Masselink, Semicond. Sci. Technol. \textbf{4}, 503 (1989).\vskip3mm
%27
\bibitem {Landau_2}L.D.~Landau and E.M.~Lifshitz, \textit{The Classical Theory
of Fields} (Pergamon Press, Oxford, 1983).\vskip3mm
%28
\bibitem {Nielsen}M.A.~Nielsen and I.L.~Chuang, \emph{Quantum Computation and
Quantum Information} (Cambridge Univ. Press, Cambridge,
2000).\vskip1mm
\begin{flushright}
{\footnotesize Received 22.04.11.\\Translated from Ukrainian by
O.I.~Voitenko}
\end{flushright}
\end{thebibliography}
\end{document}